\documentclass[%
 reprint,
 amsmath,amssymb,
prb,
showkeys,
]{revtex4-1}
\pdfoutput=1
\usepackage{graphicx}
\usepackage{dcolumn}
\usepackage{bm}
\usepackage{hyperref}

\usepackage[utf8]{inputenc}
\usepackage{rotating}
\usepackage{supertabular}
\usepackage{multirow}
\usepackage{amsfonts}
\usepackage{amsthm}
\usepackage{enumerate}
\usepackage{xcolor}
\usepackage{dsfont}
\usepackage{mathtools}
\DeclarePairedDelimiter\bra{\langle}{\rvert}
\DeclarePairedDelimiter\ket{\lvert}{\rangle}
\DeclarePairedDelimiterX\mel[2]{\langle}{\rangle}{#1 \delimsize\vert #2}

\DeclarePairedDelimiterX\Mel[2]{(}{)}{#1 \delimsize\vert #2}
\DeclareMathOperator*{\Res}{Res}
\newcommand{\BoldMath}[1]{\mathds{#1}}
\newcommand{\unit}{\BoldMath{1}}
\newcommand{\fref}[1]{Fig. \ref{#1}}

\newcommand{\MyFigure}[1]{%
 \includegraphics[width=\columnwidth]{#1}
}
\DeclareGraphicsExtensions{.pdf}

\begin{document}

\title{ Minimax Isometry Method: A compressive sensing approach for Matsubara
summation in many-body perturbation theory }

\author{Merzuk Kaltak}
 \email{merzuk.kaltak@vasp.at}
 \altaffiliation{VASP Software GmbH, Sensengasse 8/17, A-1090 Wien, Austria}
\author{Georg Kresse}
\email{georg.kresse@univie.ac.at}
\altaffiliation{University of Vienna, Faculty of Physics and Center for
Computational Materials Science,\\ 
Universit\"at Wien, Sensengasse 8/8-12, A-1090 Wien, Austria}


\date{\today}

\begin{abstract}
We present a compressive sensing approach for the long standing problem of Matsubara
summation in many-body perturbation theory. By constructing low-dimensional, almost
isometric subspaces of the Hilbert space we obtain optimum imaginary time and
frequency grids that allow for extreme data compression of fermionic and bosonic
functions in a broad temperature regime. The method is applied to the random
phase and self-consistent $GW$ approximation of the grand potential. 
Integration and transformation errors are investigated for Si and SrVO$_3$. 
\end{abstract}

\maketitle
\section{Introduction}\label{sec:Introduction}

Calculations of finite temperature properties of materials are becoming
progressively important. In particular for metals, a proper treatment of the
partial occupancies of orbitals at the Fermi level is absolutely required.  For
instance, because of a finite Brillouin zone sampling, orbitals at the Fermi level
often exhibit degeneracies and partial occupancies  that cannot be lifted without resorting to technical
tricks (such as shifting the Fermi energy).  In mean field calculations,
particularly, in density functional theory, finite temperature effects are
nowadays  usually incorporated using Mermin's
formalism,\cite{mermin1965, fu1983first, de1991ab, methfessel1989high} which has
found wide spread acceptance in most density functional theory
codes\cite{kresse1996a} and leads to a concise treatment of the grand potential,
the internal electronic energy as well as the entropy related to the electronic
degrees of freedom. Partial occupancies of degenerate states at the
Fermi level  are  thereby naturally accounted for.

For correlated wave function and Green's function based methods, including
finite temperature effects and handling partial occupancies of states at
the Fermi level is certainly less trivial but absolutely
necessary.\cite{PhysRev.118.41,PhysRev.118.1417} In Green's function theory, the
common solution is to either treat the Green's function in imaginary time and
describe it in the interval $[-\beta, \beta]$ or to impose periodicity in
imaginary time and Fourier transform all relevant quantities to imaginary
frequency.  This yields the well known  Matsubara technique,\cite{Matsubara1955}
details of which are explained in many textbooks.\cite{Fetter2003,Negele1988}

Although working in imaginary frequency and adopting the Matsubara frequencies
fundamentally allows to derive simple and compact equations for the grand
potential, internal electronic energy or the electronic entropy,
calculations using the Matsubara formulation are in many cases unpractical.
This is   especially so, if the method is combined with first principles plane  wave
codes or codes using a linear combination of atomic orbitals. For 
example, let us assume we want to calculate the properties of a
material at $T = 100$~K. This corresponds to a  Matsubara frequency spacing of about $\Delta
\omega =2 \pi /\beta = 50$~meV.  In plane wave calculations, the maximum
excitation energies are often approaching 200 to 400~eV. To perform the
required frequency summations, hence, 4000 to 8000 frequency points are
required for meV precision.  Clearly, if one were forced to use Matsubara grids,
the calculations would become intractable for all but the simplest systems and
smallest basis sets.  Thus, one has to find a way to ``compress'' the number of
grid points to an affordable small value in order to reduce the compute cost.
This is one of the main topics of the present work.  Specifically, the goal is to derive in a
mathematical concise way optimal non-uniform frequency grids that can be used
instead of the standard Matsubara grid. It goes without saying that such  grids
will always introduce small numerical errors, however, as we demonstrate in this
paper, by increasing the number of frequency points, the error drops
exponentially.  It also needs to be mentioned that these optimal grids will be
different for bosonic and fermionic functions.  This is similar to the Matsubara
technique, which results in grid points $\omega_m = (2m+1) \pi /\beta$ for
fermions and $\nu_m = 2 m \pi /\beta$ for bosons. In fact, we will see below that
this behavior is also roughly maintained for the first few frequency points for
our compressed grids. 

Another important issue is that Green's function methods can be made
particularly efficient by relying on a dual representation of all quantities in
imaginary time and imaginary frequency.\cite{Rojas1995,Rieger1999,Kaltak2015}
For instance, the well known Dyson equation $G(\omega) = G_0(\omega) +
G_0(\omega) \Sigma(\omega) G(\omega)$ is most easily solved in the frequency
domain, since the equation involves a single frequency point only. On the other
hand, the polarizability is most easily calculated in time $t$ or imaginary time
$\tau$, e.g. $\chi(t ) = -i G(t)G(-t)$ .  Thus, for an efficient implementation
it is often expedient to be able to switch via Fourier transformations from the
imaginary time to the imaginary frequency representation and vice versa without
loss of precision. Being capable to switch from imaginary frequency to
imaginary time also resolves another issue: as explained above, our compressed
grids comprise different frequencies for fermions and bosons. Hence, it is not a
simple matter to calculate a bosonic quantity from a fermionic one in frequency
space without resorting to interpolation (they are represented on different
grids). The imaginary time grid provides the necessary glue between these two grids.
The methods that we describe below adopt one and only one common grid in the
time domain (which requires a slight compromise in numerical precision). We 
also derive Fourier coefficients to bring any bosonic or fermionic function to
that common time grid. Calculations of bosonic quantities from fermionic ones are than performed in
imaginary time $\tau$, for instance $\chi(i\tau ) =  - G(-i\tau)G(+i\tau)$.

The present work is a natural extension of our own previous work on optimal zero
temperature imaginary time and frequency grids.\cite{Kaltak2014,Kaltak2015}
Furthermore, some of the ideas that we pursue here have been touched upon and are
inspired by related work published before. To name a few examples: Faleev and
coworkers used Keldysh time-loop contours to avoid explicit construction of
frequency grids.\cite{faleev2006finite}  Ku, Wei and Eguiluz suggested
non-uniform power grids to reduce the number of grid points.\cite{ku2002band}
Welden and coworkers used a Legendre representation of the imaginary time
Green's function and spline interpolation in the Matsubara
domain,\cite{Welden2016} an approach used by several other authors
before.\cite{Boehnke2011, Huang2016} All these techniques have in common that
the grids are not tailored for their purpose. Most relevant to our case is the
work of Ozaki who approximated the Fermi function by a continued fraction
representation of the hypergeometric function.\cite{Ozaki2007} Hu applied the
same method to bosons and the Bose-Einstein occupation function.\cite{Hu2010}
Shinaoka {\em et. al} developed an efficient approach for imaginary time Green's
functions using an intermediate representation between the imaginary time and
real frequency domain\cite{PhysRevB.96.035147} and Li and coworkers applied
Shinaoka's method to the $GW$ approximation recently.\cite{PhysRevB.101.035144}

The general idea of us is to map the optimization of the time and frequency
grid, or Fourier coefficients onto a well defined minimization problem. This
minimization problem is then solved using Remez's Minimax
algorithm.\cite{remez1962general, Hackbusch2008}
To obtain optimized  time and frequency grids we pursue two different routes in
the present work, corresponding to different object functions.  

The first one is
designed for non-selfconsistent perturbational many body calculations, where a
non-interacting Greens function is determined from an initial mean field
Hamiltonian. As an example, we show results for calculating the correlation
energy in the random phase approximation at finite temperature.  However, this
approach is also applicable to  M{\o}ller-Plesset perturbation theory or,
potentially, coupled cluster methods. The unifying property is that the building
blocks are always non-interacting fermionic propagators, which can be readily
obtained as resolvent of a one-particle Hamiltonian at any frequency or time
point. In this case, it suffices to solve a minimization problem that minimizes
the error in second order perturbation theory, akin to the zero temperature
case.\cite{haser:489,Kaltak2014}

Somewhat more challenging is the development of efficient sampling schemes for
self-consistent Green's function methods. In this case, the Green's function is
obtained from the Dyson equation at a set of frequency and/or time points.
This problem is more challenging, since the frequency and time grids need to be
capable to accurately represent all properties of the Green's  function {\em
and} polarization propagators without loss of the norm or spectral density.
Here, we rely on ideas previously presented by Ozaki to design optimal fermionic
Matsubara  grids that allow to represent the Fermi function with minimal
error.\cite{Ozaki2007} We, however, go beyond the work of Ozaki by mapping this
problem onto a well define minimization problem.

\section{Mathematical Formalism}\label{sec:MathematicalFormalism}

\subsection{Matsubara Technique}\label{sec:MatsubaraTechnique}
The Matsubara technique is a way to formulate quantum field theory (QFT) at
finite temperature. More precisely, it makes use of the Wick
rotation,\cite{PhysRev.96.1124} which transforms the real time axis of Minkowski
spacetime to the imaginary time axis $t\to-i\tau$. Because real space remains
unchanged by this transformation, spacetime becomes essentially euclidean, so
that this approach is also known as euclidean quantum field
theory.\cite{Osterwalder1973, Osterwalder1975}

As Matsubara has shown, the imaginary time integrals in finite temperature
perturbation theory are restricted to the interval $-\beta < \tau <
\beta$.\cite{Matsubara1955} This has the advantage that one can expand the
imaginary time-dependence of the corresponding integrands into a Fourier series,
such that imaginary-time integration becomes essentially an (infinite) series
over discrete Fourier coefficients.  The corresponding discrete frequencies are
known as {\em Matsubara frequencies} and it is important for us to distinguish
between {\em fermionic}, denoted by $\omega_n$ in the following, and {\em
bosonic} Matsubara frequencies, denoted by $\nu_m$ in the remainder of this
paper. 

Fermionic Matsubara frequencies represent the non-zero Fourier modes of
fermionic functions, while bosonic frequencies are the non-zero modes of bosonic
functions. This is explained in more detail below by means of the free-electron
Green's function (Feynman propagator) and the irreducible polarizability, the
building blocks of many-body perturbation theory.
Furthermore, if the distinction between fermionic and bosonic functions is
irrelevant we use the term {\em correlation} function.

The free propagator, or non-interacting Green's function, in imaginary time
$\tau$ represents a prototype of a fermionic function. In a one-electron basis,
the free propagator is diagonal
$g_{\alpha\gamma}(-i\tau)=\delta_{\alpha\gamma}g(x_\alpha,-i\tau)$ and the
entries read\cite{Negele1988,Fetter2003}
\begin{equation}\label{eq:DefG0}
g(x_\alpha,-i\tau)=e^{-x_\alpha\tau}
\left[
(1-f(x_\alpha))\Theta(\tau)-f(x_\alpha)\Theta(-\tau) 
\right],
\end{equation}
where $x_\alpha=\epsilon_\alpha-\mu$, and $\epsilon_\alpha$, $\mu$ and $f$ are
the one electron energy, the chemical potential and the Fermi function,
respectively.  Here $\Theta$ is the Heaviside step function.\cite{NIST} It 
is the reason why $g(x_\alpha,-i\tau)$ changes sign at $\tau=0$. Also, the
presence of the step functions implies 
\begin{equation}\label{eq:FermionicProperty}
g(x_\alpha,-i\beta+i\tau) = -g(x_\alpha,+i\tau),\quad 0<\tau< \beta.
\end{equation}
This anti-symmetric property has an important effect on the Fourier series
representation in the interval $-\beta<\tau<\beta$
\begin{align}
\label{eq:FermionicSeries}
g(x_\alpha,-i\tau)&=\frac1\beta\sum\limits_{m=-\infty}^\infty \tilde
g(x_\alpha,i\omega_m)e^{-i\omega_m \tau}\\
\label{eq:FermionicCoefficients}
\tilde g(x_\alpha,i\omega_m)&=\int_{-\frac\beta2}^{\frac\beta2} 
\mathrm{d}\tau
g(x_\alpha,-i\tau)e^{i\omega_m \tau},
\end{align}
because it contains only fermionic frequencies 
\begin{equation}\label{eq:FermionicFrequencies}
\omega_m=\frac{2m+1}{\beta}\pi,\quad m\in\BoldMath{Z}.
\end{equation}
Here and in the following, $\BoldMath{Z}$ denotes the set of all integers. 
The same representation is valid for all fermionic functions on the
imaginary time axis, including the self-energy.

\begin{figure}[tbp]
\centering
\MyFigure{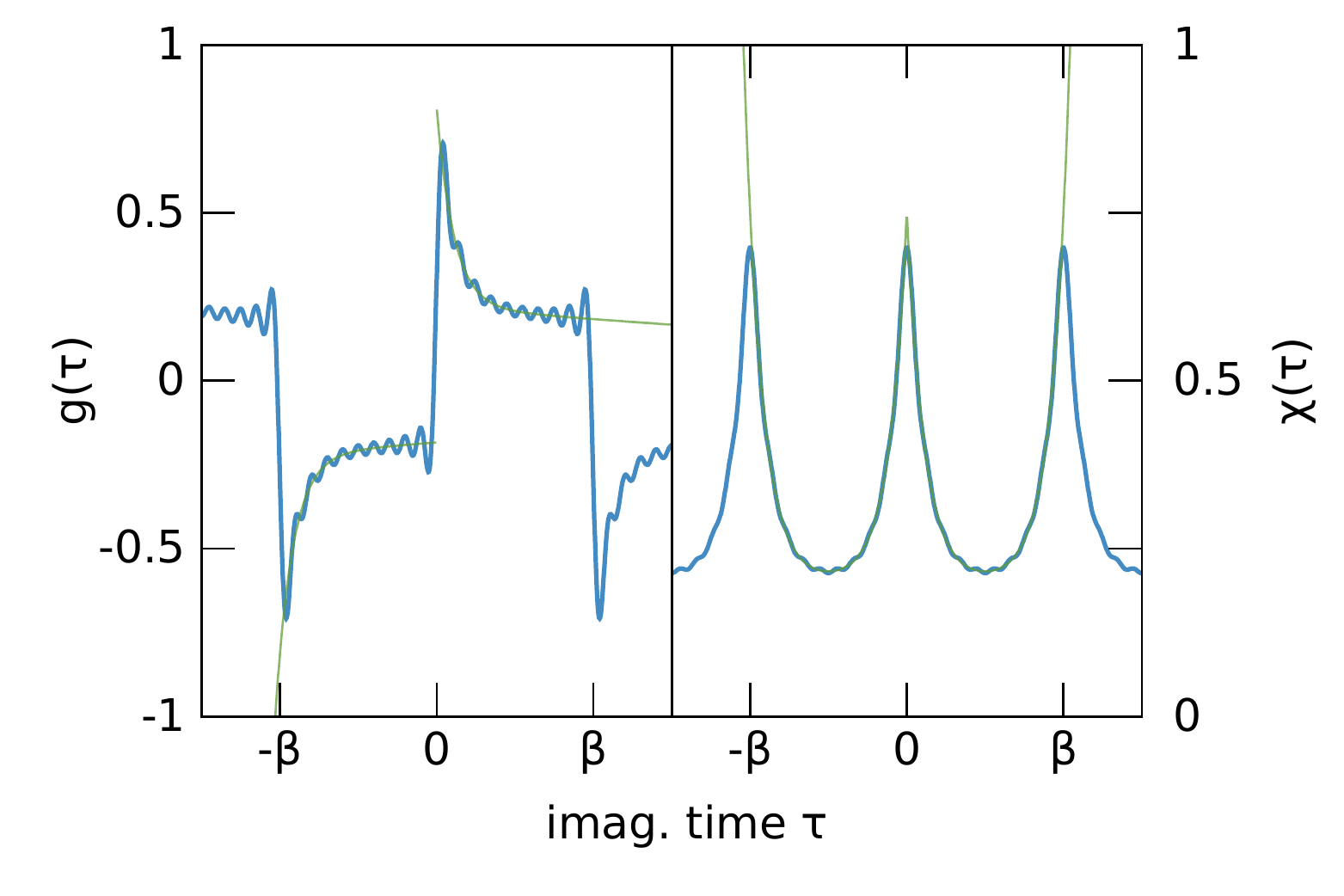}
\caption{(left) Fermionic function $g_1+g_2$ with $\epsilon_1=0.19,
\epsilon_2=9.12$ eV and $\beta=1$ eV$^{-1}$ in the fundamental interval 
(green line) and its corresponding Fourier series truncated after $m>10$ (blue
line). (right)
Corresponding bosonic function. Analytic fermionic and bosonic functions either
increase or decrease exponentially for $|\tau|>\beta$ ({\em e.g.} Equ. 
\eqref{eq:DefG0}), while the corresponding Fourier expansion outside
$[-\beta,\beta]$ is (anti-) periodic.
}
\label{fig:FermionicAndBosonicTimeDependence}
\end{figure}

An example for a bosonic function is the independent particle
polarizability, which is diagonal 
$\chi_{\alpha\gamma\alpha'\gamma'}=\delta_{\alpha\alpha'}\delta_{\gamma\gamma'}\chi_{\alpha\gamma}$
and has the entries
\begin{equation}
\label{eq:IPPolarizability}
\chi_{\alpha\gamma}(-i\tau)=-
g(x_{\alpha},-i\tau)g(x_{\gamma},+i\tau).
\end{equation}
In contrast to Equ. \eqref{eq:FermionicProperty}, bosonic functions do not
change sign in imaginary time, but are symmetric
\begin{equation}\label{eq:BosonicProperty}
\chi_{\alpha\gamma}(-i\beta+i\tau) = \chi_{\alpha\gamma}(+i\tau),\quad 0<\tau<
\beta.
\end{equation}
Consequently, the Fourier expansion 
\begin{align}
\label{eq:BosonicSeries}
\chi_{\alpha\gamma}(-i\tau)&=\frac1\beta\sum\limits_{n=-\infty}^\infty 
\tilde \chi_{\alpha\gamma}(i\nu_n)e^{-i\nu_n \tau}\\
\label{eq:BosonicCoefficients}
\tilde \chi_{\alpha\gamma}(i\nu_n)&=\int_{-\frac\beta2}^{\frac\beta2} 
\mathrm{d}\tau
\chi_{\alpha\gamma}(-i\tau)e^{i\nu_n \tau}
\end{align}
contains only bosonic frequencies
\begin{equation}\label{eq:BosonicFrequencies}
\nu_n=\frac{2n}{\beta}\pi,\quad n\in\BoldMath{Z}.
\end{equation}
It is often argued that bosonic (fermionic) functions are periodic
(anti-periodic) in $\tau$. This is strictly speaking not correct. The free
propagator, for instance,  is defined {\em a priori} only in the fundamental
imaginary time interval $|\tau|\le\beta$, because \eqref{eq:DefG0} grows or
decays exponentially for arguments outside (thin lines in
\fref{fig:FermionicAndBosonicTimeDependence}). The same holds true for the
irreducible polarizability. In fact, only the Fourier expansions
\eqref{eq:BosonicSeries} and \eqref{eq:FermionicSeries} define periodic and
anti-periodic functions in $\tau$ with (anti-) period $\beta$. This behavior is
illustrated in \fref{fig:FermionicAndBosonicTimeDependence} showing a typical
fermionic and bosonic function. In practice, it is  important to recall
that exponentially growing terms in propagators are not present, since the
$\tau$-integrations are performed over $0<\tau<\beta$ or, equivalently, over
$-\beta/2 < \tau<\beta/2$. Due to consistency with our previous
papers\cite{Kaltak2014,Kaltak2015,PhysRevB.94.165109} we usually work in the
interval $[-\beta/2,\beta/2]$.

As already explained in the introduction, the Matsubara summation has one major
drawback; the series converge very slowly with the number of frequency points,
necessitating thousands of grid points (see also Sec.
\ref{sec:FermionicFrequencyGrid}).
However, the Matsubara formalism is an elegant method to derive a closed form
for the grand canonical potential $\Omega$ of interacting
electrons.\cite{PhysRev.118.1417,Negele1988} The most important
contributions to $\Omega$ are summarized in the following section for
some commonly used approximations. 

\subsection{The correlation energy}\label{sec:RPAvsGM}

The RPA can be understood as an infinite sum of all possible ring diagrams. The method 
becomes exact for the correlation energy of the interacting homogeneous electron
gas at very high density as $T\to0$.\cite{PhysRev.106.364, PhysRev.82.625,
PhysRev.85.338, PhysRev.92.609} A closed form of the grand potential in the RPA
can be found in Negele and Orland's book\cite{Negele1988} and reads
\begin{equation}\label{eq:DefRPAGrandPotential}
\Omega_c^{\rm RPA} = \frac12
\frac1\beta\sum\limits_{n\in\BoldMath{Z}}\mathrm{Tr}\left\lbrace
\ln\left[\unit-\tilde{\chi}(i\nu_n)V\right] 
-\tilde\chi(i\nu_m)V
\right\rbrace,
\end{equation}
where $V$ stands for the Coulomb matrix elements and the trace $\mathrm{Tr}$ refers to summation
over elements of the basis.  In second order this corresponds
to the direct term in M{\o}ller-Plessett (MP2) perturbation theory:
\begin{equation}\label{eq:DefMP2GrandPotential}
\begin{split}
\Omega_c^{\rm d-MP2} = &\frac14
\frac1\beta\sum\limits_{n\in\BoldMath{Z}}\mathrm{Tr}\left[
\tilde{\chi}(i\nu_n)V
\tilde{\chi}(i\nu_n)V 
\right] \\
=&\frac14
\int_{-\frac\beta2}^{\frac\beta2}\mathrm{d}\tau\mathrm{Tr}\left[
\chi(-i\tau)V
\chi(i\tau)V
\right].
\end{split}
\end{equation}
A key point is that the correlation energy in both, the RPA and MP2, depends only
on the polarizability, respectively, on products of two Green's functions
$G(i\tau) G(-i\tau)$. In this sense, the RPA is an approximative bosonization of
the original problem, a property that greatly simplifies the construction of
appropriate time and frequency grids.  Specifically, only the bosonic
frequencies $\nu_n$ enter in the final evaluation of the correlation energy.

As an example of methods where bosonization is typically not applicable, we
decided  to evaluate the Galitskii-Migdal (GM)
expression\cite{GalitskiiMigdal1957, PhysRevLett.110.146403} for the correlation
part of the grand potential
\begin{equation}\label{eq:DefGMGrandPotential}
\begin{split}
\Omega_c^{\rm GM} =&
\frac1\beta\sum\limits_{m\in\BoldMath{Z}} \mathrm{Tr}\left[
\tilde{G}(i\omega_m)\tilde{\Sigma}(i\omega_m)
\right]\\
=&
\int_{-\frac\beta2}^{\frac\beta2}\mathrm{d}\tau \mathrm{Tr}\left[
G(i\tau)\Sigma(-i\tau)
\right].
\end{split}
\end{equation}
Here $\tilde{G}$ is the dressed propagator and the solution of 
the Dyson equation
\begin{equation}\label{eq:DysonG}
\tilde{G}(i\omega_m) = 
\tilde{G}_0(i\omega_m) +
\tilde{G}_0(i\omega_m)
\tilde{\Sigma}(i\omega_m)
\tilde{G}(i\omega_m),
\end{equation}
where $\tilde G_0$ is the Hartree-Fock Green's function, $\tilde{\Sigma}$ the
$GW$ correlation self-energy\cite{Hedin1965} 
\begin{equation}\label{eq:SelfEnergyGW}
\tilde{\Sigma}(i\omega_m) = \int_{-\beta/2}^{\beta/2} 
\mathrm{d}\tau
G(-i\tau)W(-i\tau)e^{i \omega_m \tau}
\end{equation}
and $\tilde{W}$ the RPA screened potential 
\begin{equation}\label{eq:GWPotentialW}
\tilde{W}(i\nu_n) = 
V +
V
\tilde{\chi}(i\nu_n)
\tilde{W}(i\nu_m).
\end{equation}
Considering the equations above, it should be quite obvious why it is
substantially more difficult to obtain suitable time and frequency grids in this
case. Equ. (\ref{eq:GWPotentialW}) should be solved on the bosonic frequency
grid, whereas all the other quantities need to be evaluated on fermionic grids.
This implies that at least two frequency grids (and potentially time grids) are required.

\subsection{Odd and even functions of time}\label{sec:oddandeven}
We found it expedient to distinguish between the time-symmetric and
anti-symmetric part of the Green's function
\begin{align}
\label{eq:SymmetricPart}
\hat u_\tau(x_\alpha)=&\frac{g(x_\alpha,-i\tau)+g(x_\alpha,+i\tau)}2\\
\label{eq:AntisymmetricPart}
\hat v_\tau(x_\alpha)=&\frac{g(x_\alpha,-i\tau)-g(x_\alpha,+i\tau)}2.
\end{align}
It is easy to show  that for the independent particle Green's function \eqref{eq:DefG0} in the
interval $[-\beta,\beta]$ the corresponding even and odd functions read
\begin{align}
\label{eq:SymmetricBasis}
\hat u_\tau(x_\alpha)=&\frac12\frac{
\sinh\left[\frac{\beta x_\alpha}{2}
\left(1-2\frac{\left|\tau\right|}{\beta}\right)\right]
}{
\cosh\left(\frac{\beta x_\alpha}{2}\right)
}\\
\label{eq:AntisymmetricBasis}
\hat v_\tau(x_\alpha)=&\frac{\mathrm{sgn}(\tau)}2\frac{
\cosh\left[\frac{\beta x_\alpha}{2}
\left(1-2\frac{\left|\tau\right|}{\beta}\right)\right]
}{
\cosh\left(\frac{\beta x_\alpha}{2}\right)
}
\end{align}
and have the following Fourier coefficient functions at fermionic
Matsubara frequencies $\omega_n$
\begin{align}
\label{eq:SymmetricPartG0InFreq}
\tilde u_{\omega_n}(x_\alpha)=&
\frac12\frac{x_\alpha}{x_\alpha^2+\omega_n^2}
\\
\label{eq:AntisymmetricPartG0InFreq}
\tilde v_{\omega_n}(x_\alpha)=&
\frac12\frac{\omega_n}{x_\alpha^2+\omega_n^2}.
\end{align}
If we generalize the above functions to bosonic functions by defining
\begin{align}
\label{eq:SymmetricBasisBoson}
 u_\tau(x_\alpha)=&\frac12 \frac{
\cosh\left[\frac{\beta x_\alpha}{2}
\left(1-2\frac{\left|\tau\right|}{\beta}\right)\right]
}{
\cosh\left(\frac{\beta x_\alpha}{2}\right)
}
\\
\label{eq:AntisymmetricBasisBoson0}
 v_\tau(x_\alpha)=&  \frac{\mathrm{sgn}(\tau)}2 \frac{
\sinh\left[\frac{\beta x_\alpha}{2}
\left(1-2\frac{\left|\tau\right|}{\beta}\right)\right]
}{
\cosh\left(\frac{\beta x_\alpha}{2}\right),
}
\end{align}
we obtain the corresponding bosonic Fourier coefficient functions at bosonic
frequencies $\nu_n$ as
\begin{align}
\overline u_{\nu_n}(x_\alpha)=&
\frac12\frac{x_\alpha}{x_\alpha^2+\nu_n^2}\tanh\frac{ x_\alpha \beta }{2}
\label{eq:SymmetricPartChiInFreq}
  \\
  \label{eq:AntisymmetricPartChiInFreq}
\overline v_{\nu_n}(x_\alpha)=&
\frac12\frac{\nu_n}{x_\alpha^2+\nu_n^2}\tanh\frac{ x_\alpha \beta }{2}.
\end{align}
Table \ref{tab:IsometricBases} summarizes the functions defined in this manner.
Note that the bosonic and fermionic function are identical in time, if we
restrict  the value of $\tau$  to $[0,\beta/2]$.  An advantage of defining odd
and even functions is that one can restrict all time  integrations to the
interval  $[0,\beta/2]$ and obtain the results from negative imaginary times by
symmetry considerations. Also, summations over Matsubara frequencies can be
constrained to positive frequencies, since the contributions from negative
frequencies follow again from symmetry considerations. 

The basis functions defined above reduce to our previously used
basis functions in the $\beta\to\infty$ limit.\cite{Kaltak2014,
PhysRevB.94.165109} More precisely, the even and odd imaginary time basis
functions \eqref{eq:SymmetricBasis}, \eqref{eq:AntisymmetricBasis},
\eqref{eq:SymmetricBasisBoson}, \eqref{eq:AntisymmetricBasisBoson0} approach all the
same limit on the positive $\tau$-axis for $\beta\to\infty$, namely the zero
temperature basis function $\frac12 e^{-|x_\alpha\tau|}$.  The
Fourier bases \eqref{eq:SymmetricPartChiInFreq},
\eqref{eq:AntisymmetricPartChiInFreq} and \eqref{eq:SymmetricPartG0InFreq},
\eqref{eq:AntisymmetricPartG0InFreq} separate into two distinct basis functions
in this limit (see IC in Tab. \ref{tab:IsometricBases}).  This means that
for $\beta\to\infty$ there is only one optimal $\tau$-grid and two distinct
optimal frequency grids for the functions defined above; a fact that has been
exploited by the authors in previous papers.\cite{Kaltak2015,
PhysRevB.94.165109, PhysRevB.98.155143} 

The {\em duality principle} between time and frequency, which was formulated in
our previous papers, allows transformations between grid representations of the
same quantity without significant loss in precision. In the present work, it  is
understood rigorously in terms of {\em almost isometric spaces} discussed in the
next section.  We employ this method to derive a compressed representation of
the independent-particle polarizability at finite temperature that allows for
accurate summations over bosonic Matsubara frequencies in Section
\ref{sec:TimeFrequencyRPA}.

To this end,  we prove a general theorem about almost isometric Hilbert spaces
that can be used to determine compressed representations for the polarizability
in imaginary time and imaginary frequency.  The corresponding time and frequency
grids are ideal to calculate the RPA correlation energy at finite
temperature with a small number of grid points. 

\begin{table}
\caption{Almost isometric basis functions $\mel*{\omega}x$ and $\mel*{\tau}x$
related by cosine (subscript 1) or sine transformations (subscript 2) for
$\beta=1$.  IA$_1$ and IB$_2$ represent bosonic (b) functions, whereas  IA$_2$
and IB$_1$ represent fermionic (f) functions.  $\mel*{\omega}x$ must be
evaluated at the respective Matsubara grids.  Third column shows the
corresponding conserved $L^2$-norm for $N\to\infty$ (given by Equ.
\eqref{eq:Conservation}).  From the infinite set of basis functions a discrete
set with time points $\lbrace \tau_j^*\rbrace_{j=1}^N$ and frequency points
$\lbrace \omega_k^*\rbrace_{k=1}^N$ will be selected to independently minimize
the errors in the $L^2$-norm tabulated in the column $\|x\|_2^2$.  IB$_2$ is not
relevant for the present work, since polarization propagators observe the
symmetries IA$_1$.  }

\begin{tabular*}{0.49\textwidth}{@{\extracolsep{\fill}}ccccc}
\hline
\hline
group  &   &$\mel*{\tau}x$ &  $\mel*{\omega}x$ &  $\| x\|_2^2$    \\
\hline                      
\\
 IA$_1$ & b
& $\frac12\frac{\cosh\frac{x}{2}(1-2|\tau|)}{\cosh\frac{x}{2}}$ 
& $\frac12\frac{x\tanh\frac{x}{2}}{x^2+\omega^2}$                   
&\multirow{2}{*}{ $\frac{\tanh\frac{x}{2}}{4x}+\frac{1-\tanh^2\frac{x}{2}}{8}$ }
\\
 IA$_2$ & f   
& $\frac{\mathrm{sgn}(\tau)}{2}\frac{\cosh\frac{x}{2}(1-2|\tau|)}{\cosh\frac{x}{2}}$ 
& $\frac12\frac{\omega}{x^2+\omega^2}$                                
\\
\\
\hline
\\
 IB$_1$ & f 
&$\frac12\frac{\sinh\frac{x}{2}(1-2|\tau|)}{\cosh\frac{x}{2}}$ 
& $\frac12\frac{x}{x^2+\omega^2}$                
&\multirow{2}{*}{ $\frac{\tanh\frac{x}{2}}{4x}-\frac{1-\tanh^2\frac{x}{2}}{8}$ }
\\
 $[$IB$_2$$]$ & b  
&$\frac{\mathrm{sgn}(\tau)}{2}\frac{\sinh\frac{x}{2}(1-2|\tau|)}{\cosh\frac{x}{2}}$ 
&$\frac12\frac{\omega\tanh\frac{x}{2}}{x^2+\omega^2}$                
& 
\\
\\
\hline
\\
 IC$_1$ & b 
& $\frac12 e^{-|x\tau|} $ 
& $\frac12\frac{|x|}{x^2+\omega^2}$                
&\multirow{2}{*}{ $\frac1{4x}$}
\\
 IC$_2$ & f 
& $ \frac{\mathrm{sgn}(\tau)}{2}e^{-|x\tau|} $ 
& $\frac12\frac{\omega}{x^2+\omega^2}$                
& 
\\
\\
\hline
\hline
\end{tabular*}
\label{tab:IsometricBases}
\end{table}

\begin{figure}[tbp]
\centering
\MyFigure{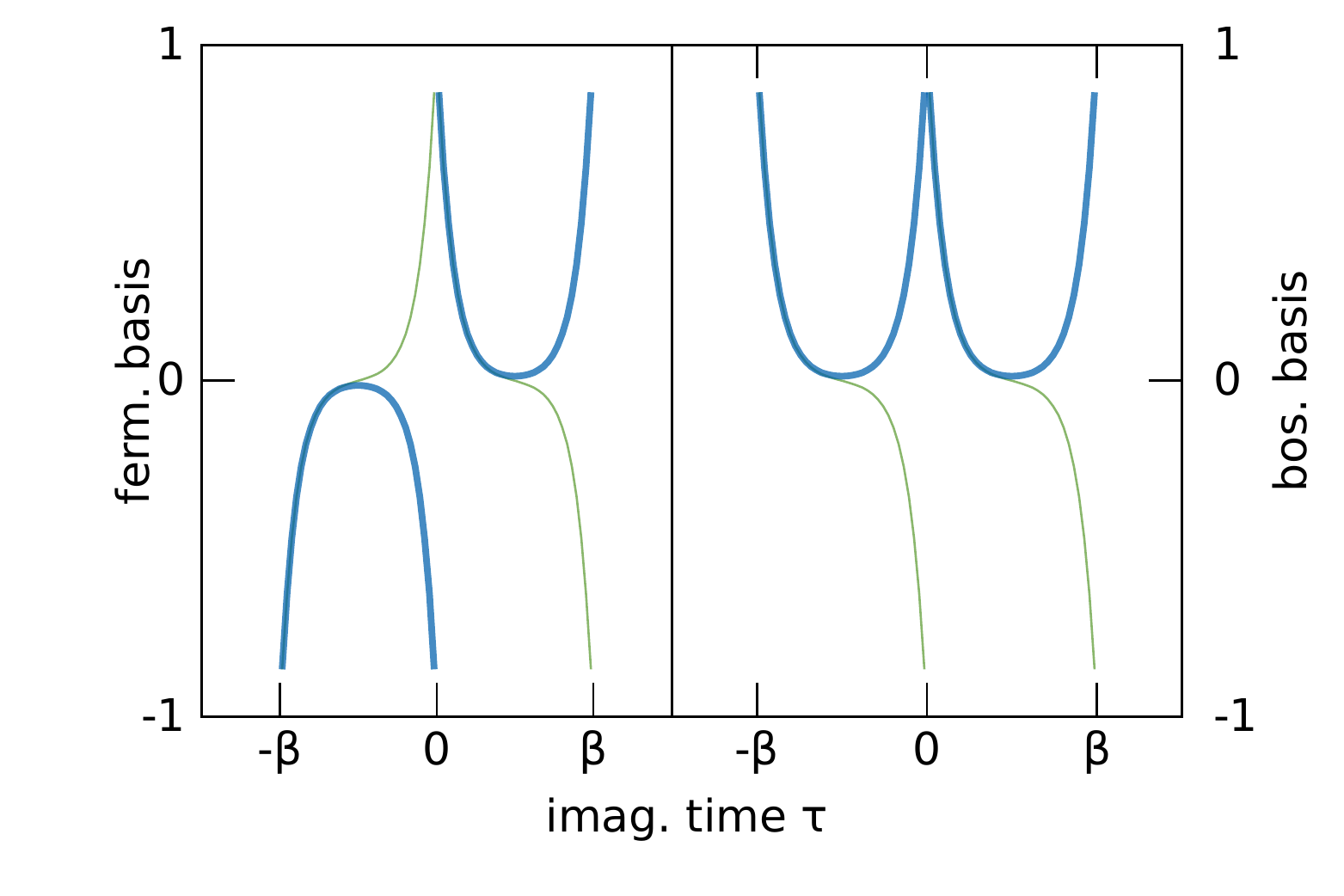}
\caption{(left) Fermionic basis functions $\mel*{\tau}{x}$ IA$_2$ (thick line)
and IB$_1$ (thin line) in imaginary time for $x=10$.  (right) Corresponding
bosonic basis functions IA$_1$ (thick) and IB$_2$ (thin). }
\label{fig:BasisFunctions}
\end{figure}

For fermionic functions, the situation is more complicated, because
the Green's function $\tilde G$ or the self-energy $\tilde \Sigma$ can be
represented only using both basis functions \eqref{eq:SymmetricPartG0InFreq} and
\eqref{eq:AntisymmetricPartG0InFreq}. How to obtain a compressed fermionic
frequency grid that describes both basis functions accurately is discussed in section
\ref{sec:FrequencyGM}.

\section{Time and Frequency Grids for RPA}\label{sec:TimeFrequencyRPA}

Obviously, for every one-electron energy $x_\alpha$, one obtains a corresponding
contribution to the Green's functions $\hat v_\tau(x_\alpha), \hat
u_\tau(x_\alpha)$ and $\tilde v_\omega(x_\alpha), \tilde  u_\omega(x_\alpha)$,
respectively. Likewise, for a typical transition energy $x_\alpha=\epsilon_{\rm
unocc}- \epsilon_{\rm occ}$, one obtains contributions to the independent
particle polarizability $\chi$ following approximately $v_\tau(x_\alpha),
u_\tau(x_\alpha)$  and $\overline v_\omega(x_\alpha), \overline
u_\omega(x_\alpha)$ in time and frequency.  The relevant question is whether one
can chose an {\em optimal discrete set of frequencies $\omega_k^*$ and time
points $\tau_k^*$} $k=1,...,N$ with corresponding functions that allow  to
represent all possible contributions to the Green's function and polarizability
accurately.  Importantly,  in the next subsection we will drop the constraint
that the frequencies must correspond to Matsubara frequencies, but we maintain
the functional form for the frequency dependence. This is a key step of the
present approach. 

\subsection{Minimax Isometry}\label{sec:MinimaxIsometry}
To keep the notation simple, we consider the case $\beta=1$ eV$^{-1}$ in the
following. The general case $\beta\neq1$ follows from scaling relations that are
discussed in \ref{sec:TimeGrid} and \ref{sec:BosonicFrequencyGrid}.  The energy
levels and transition energies are supposed to be bound $x\in[0,x_{\rm max}]$,
as is typically the case in first principles calculations. That is, the interval
length $x_{\rm max}$ is either the largest eigenenergy $(\epsilon_{\rm max}-\mu)
\beta$ or the largest transition energy $(\epsilon_{\rm max}-\epsilon_{\rm
min})\beta$.  Furthermore, we consider the functions $v_\tau(x),
u_\tau(x),\tilde u_\omega(x),\cdots$ as the time and frequency representations
of an abstract vector $\ket*{x}$ in a Hilbert space $\mathcal{H}$ and define the
in-products $\langle \tau \ket*{x}$ as its imaginary time and $\langle \omega
\ket*{x}$ as its imaginary frequency representation, which are equivalent to
the corresponding functions in Tab. \ref{tab:IsometricBases}.

It is assumed that $\ket*{\tau}$ and $\ket*{n}$ (shorthand for $\ket{\omega_n}$
or $\ket{\nu_n}$) are two complete basis sets in imaginary time and frequency
for the same function space, such that the identity operator $\unit$ can be
expressed as 
\begin{align}
\label{eq:Completeness1}
\unit=&\int_0^{1/2}\mathrm{d}\tau \ket*{\tau}\bra*{\tau}\\
\label{eq:Completeness2}
\unit=& \sum_{n\in\BoldMath{Z}}\ket*{n}\bra*{n}.
\end{align}
From a functional analysis perspective, one says that the two spaces
$\mathcal{X}=\mathrm{span}\lbrace \ket*\tau\rbrace_{\tau\in[0,1/2]}$ and
$\tilde{\mathcal{X}}=\mathrm{span}\lbrace\ket*{n}\rbrace_{n\in\BoldMath{Z}}$ are
isometric with respect to the scalar product induced norm $\|x\|_2=\sqrt{\mel*{x}x}$
so that
$(\mathcal{X},\mel*{\cdot}{\cdot})\cong(\tilde{\mathcal{X}},\mel*{\cdot}{\cdot})$.
This isometry (indicated by the symbol $\cong$) is effectively a simple basis
transformation that does not change the induced norm, since
\begin{equation}\label{eq:Conservation}
\begin{split}
\|x\|_2^2 = \mel*{x}{x} = &
\int_0^{1/2}\mathrm{d}\tau\mel*{x}\tau\mel*\tau{x}\\
=&
\sum_{n\in\BoldMath{Z}}\mel*{x}n\mel*{n}{x}. 
\end{split}
\end{equation}
It is a simple matter, to show that both the time integral 
as well as the frequency summation in Equ. (\ref{eq:Conservation})
indeed yield the same result, which are shown in the final column in Tab.
\ref{tab:IsometricBases} (this shows that the two basis sets are indeed
isometric).\footnote{ The isometry \eqref{eq:Conservation} is known
as Parseval theorem\cite{Boto2013} or, if $\ket*{n}$ is continuous,
Plancherel theorem for Fourier transforms.\cite{Plancherel1910}.}

If $\ket*\tau$ is the time and $\ket*{n}$ the discrete
 frequency basis for fermions (bosons), then
$\mel*\tau{n}$ and $\mel*{n}\tau$ are the matrix elements $\cos(\omega_n
\tau),\sin(\omega_n\tau),\cos(\nu_n\tau)$ or $\sin(\nu_n\tau)$ of the forward
and backward basis transformation.  Consequently, the two spaces $\mathcal{X}$
and $\tilde{\mathcal{X}}$ are equivalent and span the same Hilbert space $\mathcal{H}$.
This equivalence holds true only if infinitely many basis vectors are
considered; for finite dimensional subspaces the perfect isometry
\eqref{eq:Conservation} is violated. 

One may illustrate the violation of the isometry \eqref{eq:Conservation} with
the discrete Fourier transform (DFT) having the bases
$\ket*{\tau_k}=\ket*{\frac{1}{2N}(2k-1)}$ with $(k=1,\cdots,N)$ and
$\ket*{n}=\ket{\pi(2n-1)}$ (truncated fermionic Matsubara grid). The
corresponding completeness relations \eqref{eq:Completeness1},
\eqref{eq:Completeness2} become projectors onto finite dimensional subspaces
$X\subset\mathcal{X},\tilde{X}\subset\tilde{\mathcal{X}}$ and have the form
\begin{align}
\label{eq:DFTCompleteness1}
P=&\frac1{2N}\sum\limits_{k=1}^N\ket*{\tau_k}\bra*{\tau_k}\\
\label{eq:DFTCompleteness2}
\tilde{P}=& \sum_{n=-N+1}^N\ket*{n}\bra*{n}=2\sum_{n=1}^N\ket*{n}\bra*{n}
\end{align}
Only in the limit $N\to\infty$ the projectors approach the identity operator
$\unit$. For finite $N$, the isometry \eqref{eq:Conservation} is violated, but
can be replaced by a so-called
$\varepsilon$-isometry\cite{Fleming2002,Ding1988}
\begin{equation}\label{eq:EpsilonIsometry} 
\| P - \tilde P\|:=\max_{0\le x\le x_{\rm max}}|\bra*{x} P - \tilde P \ket*{x}| \le
\varepsilon.
\end{equation}
Of interest to us is the magnitude of $\varepsilon$ and especially how it
decreases with increasing $N$.  For instance, in the case of the DFT
$\bra*{x}P\ket*{x}$ is the Riemann sum of the integral in
\eqref{eq:Conservation} of order $N$ and is known to be a poor method to
evaluate integrals. As a consequence, $\varepsilon$ of the Matsubara grid is a
weakly decaying function in $N$ and cannot be used for our purposes, as shown in
section \ref{sec:MinimaxFourierTransformation}.

The following question naturally arises: how can one determine 
$\varepsilon$-isometric subspaces $X=\mathrm{span}\lbrace\ket*{\tau_k}\rbrace_{k=1}^N$,
and $\tilde{X}=\mathrm{span}\lbrace\ket*{\omega_k}\rbrace_{k=1}^N$,
such that the completeness relations \eqref{eq:Completeness1} and
\eqref{eq:Completeness2} are approximated as good as possible for all vectors
$\ket*{x}$ with $0\le x\le x_{\rm max}$? 

Using the notation in \eqref{eq:EpsilonIsometry}, the answer to this question
are the solutions of following minimax problems:
\begin{align}
\label{eq:MMU}
\min_{\sigma_k>0,\tau_k\in(0,1/2)} &
\left\| 
\unit - \sum\limits_{k=1}^N\sigma_k \ket*{\tau_k}\bra*{\tau_k} 
\right\|
\\
\label{eq:MMV}
\min_{\lambda_k>0,\omega_k>0} &
\left\| 
\unit - \sum\limits_{k=1}^N\lambda_k \ket*{\omega_k}\bra*{\omega_k} 
\right\|.
\end{align}
Provided the solutions exist, they are known to yield errors $\varepsilon$ that
decay exponentially with $N$.\cite{Braess1986} In the following, we prove that
\eqref{eq:MMU} and \eqref{eq:MMV} satisfy our requirements. 

To prove the assertion above it suffices to show that the minimax errors are an
upper bound for the isometry violation in \eqref{eq:EpsilonIsometry}.
Therefore, assume $\lbrace\sigma_k^*,\tau_k^*\rbrace_{k=1}^N$ and
$\lbrace\lambda^*_k,\omega_k^*\rbrace_{k=1}^N$ are the solutions of
\eqref{eq:MMU} and \eqref{eq:MMV} with
$P^*=\sum_{k=1}^N\sigma^*_k\ket*{\tau_k^*}\bra*{\tau_k^*}$ and $\tilde
P^*=\sum_{k=1}^N\lambda^*_k\ket*{\omega_k^*}\bra*{\omega_k^*}$ the corresponding
projectors, respectively. Then a positive number $\varepsilon/2$  exists (for
every given $N$) as an upper bound for \eqref{eq:MMU} and \eqref{eq:MMV} and one
can write
\begin{equation}
\label{eq:ProofEq1}
\begin{split}
\Big\| 
\unit - P^*
\Big\|
 \le &\frac12 \varepsilon\\
\left\| 
\unit - \tilde P^*
\right\| = 
\left\| 
\tilde P^*-\unit
\right\| 
 \le &\frac12 \varepsilon.
\end{split}
\end{equation}
Adding both inequalities in \eqref{eq:ProofEq1} and using the triangle
inequality $ \|f+g\|\le\|f\|+\|g\|$ (satisfied by every norm\cite{Boto2013}) one 
obtains 
\begin{equation}
\label{eq:ProofEq2}
\begin{split}
\underbrace{
\Big\| 
\unit-P^*
\Big{\|}
 +
\left\| 
\tilde P^*-\unit
\right\|
}_{
\left\| 
\unit-P^*+\tilde P^*-\unit
\right\| 
\le
}  \le \varepsilon.
\end{split}
\end{equation}
Last inequality implies \eqref{eq:EpsilonIsometry} for the projectors
$P^*,\tilde P^*$ and concludes our proof.  \qed

This is a quite remarkable result, because it means that the projectors $P^*$ and
$\tilde P^*$ converge to the identity operator and, thus,  define
$\varepsilon$-isometric topological vector spaces $X^*,\tilde{X}^*$ that have the {\em
approximation property}.\cite{Schaefer1999} A summary of $\varepsilon$-isometric
bases is given in Tab.  \ref{tab:IsometricBases} and discussed
below.

Note that the discussion above does not give a prescription how to determine the
transformation $X^*\to \tilde{X}^*$; a corresponding method is presented in section
\ref{sec:MinimaxFourierTransformation}.

The proof above contains only an upper bound for the transformation error in
\eqref{eq:ProofEq2}. This upper bound $\varepsilon$ is inherited from the sum of
the convergence rate of the minimax solutions in the $\tau$- and
$\omega$-domain. This convergence rate has been studied by Braess and Hackbusch
for the minimax problem IC in the $\tau$-domain listed in Tab.
\ref{tab:IsometricBases}. They obtained
$\varepsilon(N)\approx6.7\log(2+N)e^{-\pi\sqrt{2N}}$ for $x\in [1,R_N]$, where
$[1,R_N]$ belongs to the largest possible error for a given order
$N$.\cite{Braess2005} Our numerical experiments discussed in section
\ref{sec:Results} indicate similar convergence rates for all other minimax
problems in Tab.  \ref{tab:IsometricBases}.  In contrast, the DFT or Matsubara
grid has only a linear rate of convergence $\varepsilon(N)\propto N^{-1}$. 

\subsection{Discussion of Isometry}\label{sec:BasisFunctions}

In this subsection, we try to give more insight into what we have achieved at
this point.  We start with the IC basis, which has been used in previous
publications by the authors to construct optimized minimax grids for low scaling
random phase and $GW$ algorithms at zero temperature using a different line of
arguments.\cite{Kaltak2014,Kaltak2015,PhysRevB.94.165109} Specifically,
$\mel*{\tau}x$ of IC$_1$ describes the imaginary time dependence of the
independent particle polarizability at zero temperature for a transition energy
$x$, while the corresponding $\mel*{\nu}x$ functions describe its imaginary
frequency dependence.\cite{Kaltak2014} The corresponding conserved $L^2$-norm
(forth column of Tab. \ref{tab:IsometricBases}) is the key quantity for the
second order contributions to the correlation energy (see Equ.
\eqref{eq:DefMP2GrandPotential} or Ref. \onlinecite{Kaltak2014,Hackbusch2008}).
These contributions involve energy denominators of the form $1/( 2\epsilon_a-
2\epsilon_i)$ and are considered to be bound, i.e. virtual states with
energies $\epsilon_a$ are separated by a band gap $\Delta_{ai}$ from occupied
states with energy $\epsilon_i$.\cite{Kaltak2014}  The induced norm is important
and essentially tells the optimization in the minimax problem which
contributions to the energy are most relevant. With this choice,  contributions
from small energy differences dominate over contributions from large energy
differences, hence in the frequency space the grid points will be more densely
spaced at small frequencies.

The $\tau$-basis function of IC$_2$ has the same time dependence (apart from the
opposite sign on the negative $\tau$-axis), while the imaginary frequency
dependence of the cosine transformation differs considerably from the one
obtained from the sine transform (compare $\mel*{\omega}x$ of IC$_1$ and
IC$_2$).  It comes with no surprise that the  minimax frequency grids for both
are different too.\cite{PhysRevB.94.165109} However, the time grids are
identical, and the minimax isometry guarantees that one can map in time between
IC$_1$ and IC$_2$ with high precision. 

Next, we consider the four basis functions of group IA and IB. They can be
grouped into bosonic (IA$_1$ and IB$_2$) and fermionic (IA$_2$ and IB$_1$)
pairs. When optimizing the frequency grid points using the minimax algorithm, we allow
$\omega$ in IA and IB to deviate from the corresponding Matsubara grid. Indeed,
the corresponding Minimax solutions are non-uniformly distributed, but
nevertheless closely match Matsubara frequencies at small $\omega$.  It turns
out, as shown in section \ref{sec:Results}, that this freedom allows us to
describe the high frequency tail of the correlation functions with high
precision even in low dimensional subspaces $X^*, \tilde{X}^*$ without the need for
interpolation.  The corresponding $\varepsilon$-isometric time basis functions
have the fermionic anti-symmetry [Equ.  \eqref{eq:FermionicProperty}] and
bosonic symmetry [Equ.  \eqref{eq:BosonicProperty}]  for $-\beta/2\le
\tau\le\beta/2$, respectively, and are illustrated in
\fref{fig:BasisFunctions}.

At finite temperature, the situation is analogous to the zero temperature case,
i.e. the conserved $L^2$-norm of the IA isometry describes the second order
contribution to the correlation part of the grand-canonical potential defined in
Equ.  \eqref{eq:DefMP2GrandPotential} (see appendix \ref{app:Proof2ndorder}).
Because of time-inversion symmetry, the polarizability, the screened potential,
or contributions to the correlation energy can be entirely presented  by IA$_1$
basis functions since e.g. $\chi(-i\tau)=\chi(-i\beta+i\tau)$ (blue lines in
\fref{fig:BasisFunctions} right).  Thus, the isometry IA$_1$ can be employed to
obtain compressed time and frequency grids for the calculation of the correlation
part of the grand canonical potential in the RPA as well as MP2.  The
corresponding imaginary time and frequency grid are discussed in
\ref{sec:TimeGrid} and \ref{sec:BosonicFrequencyGrid}, respectively. 

As already emphasized before, if we use self-consistent techniques and the GM
formula for the grand canonical potential the construction of optimal time and
especially frequency grids becomes more difficult, since even {\em and} odd
basis functions in the frequency domain (IB$_1$ and IA$_2$) contribute to the
grand potential and have different $L^2$-norms (compare third column of IA and
IB).  We, therefore, propose an alternative approach in this case that is based
on the minimization of the $L^1$-quadrature error instead (see section
\ref{sec:FermionicFrequencyGrid}).

\subsection{Imaginary time grid}\label{sec:TimeGrid}
To construct an imaginary time grid for arbitrary $\beta$, we make use of the
scaling properties
\begin{equation}\label{eq:ScalingTime} 
\tau_j \to \beta \tau_j, \quad \sigma_j
\to \beta \sigma_j 
\end{equation} 
that allow to recover the time quadrature for an arbitrary interval $[0,\beta/2]$
from the unscaled solution determined for $[0,1/2]$.

How to chose the bosonic time grid (IA$_1$) has been discussed in the previous
section.  However, we also need a time grid to represent  fermionic quantities,
such as the Green's functions from which the polarizabilities are build as $g(-i
\tau)g(+i \tau)$.  For computational reasons, it  is obviously desireable  to
use only one time grid, since this allows us to represent the Green's functions
and the bosonic quantities on the same time grid.  The even and odd basis
functions of the IA and IB $\varepsilon$-isometry in Tab.
\ref{tab:IsometricBases} are clearly identical for fermions and bosons at
positive $\tau$,
\begin{align}\label{eq:BasisTimeEven}
u_\tau(x)=&\frac12\frac{\cosh\frac{x}{2}(1-2\tau)}{\cosh\frac{x}{2}},\quad
\tau>0\\
\label{eq:BasisTimeOdd}
v_\tau(x)=&\frac12\frac{\sinh\frac{x}{2}(1-2\tau)}{\cosh\frac{x}{2}},\quad
\tau>0.
\end{align}
Odd functions are not relevant for bosons as argued above, however, they do matter
for fermions, and  optimization of the time grid for even and odd functions yields
different time grids.  We opt to 
use the optimal even time grid (IA$_1$) as a common grid for both
fermionic and bosonic functions and summarize the relevant arguments here.  (i) The second
order and RPA correlation energy depends only on bosonic functions, e.g. the
polarizability. Hence the fermionic functions are only used at an intermediate
stage. (ii) The imaginary time grid for the even
functions $u$ yields a small minimax error also for the odd basis functions $v$
for the entire interval $x\in[0,x_{\rm max}]$, with larger but still negligible errors even for $x\to
0$. We suspect that this is due to the fact that in the zero temperature limit
$\beta\to\infty$ both basis functions \eqref{eq:BasisTimeEven} and
\eqref{eq:BasisTimeOdd} approach smoothly the same exponential form in the interval
$\tau\in[0,1/2]$ (see IC isometry in Tab. \ref{tab:IsometricBases}).

In summary, we solve the minimax problem \eqref{eq:MMU} only for IA$_1$
$\mel*{\tau_j}{x}= u_{\tau_j}(x)=:u_j(x)$ and use the same time grid points
$\tau_j$ for the odd fermionic basis functions.
To obtain the minimax time grid points $\tau^*_j$, it is convenient to rewrite the minimax
problem \eqref{eq:MMU} into the following form
\begin{equation}\label{eq:MinimaxProblemTime}
\min_{\sigma_j>0,\tau_j\in(0,1/2)}
\max_{0\le x\le x_{\rm \max}}
\left|
\|x\|_2^2
-\sum\limits_{j=1}^{N}\sigma_j
 u_j^2(x)
\right|
\end{equation}
with $x_{\rm max}=\beta\epsilon_{\max}$ and
$\epsilon_{\max}$ the maximum one-electron energy considered. Then it becomes
evident that \eqref{eq:MinimaxProblemTime} is a non-linear fitting problem of
separable type,\cite{Golub2003} which in general has only a solution, if every
basis function $ u_j$ is linearly independent and has less than $N-1$
zeros.
The alternant theorem then implies\cite{Braess1986, Haemmerlin1994} a set of
points $\lbrace x^*_k\rbrace_{k=0}^{2N}$ (alternant) and a set of non-linear
equations
\begin{equation}\label{eq:AlternantTheorem}
\|x_k^*\|_2^2-\sum\limits_{j=1}^N\sigma^*_ju^2_{\tau_j^*}(x_k^*)=(-1)^k E_N
\end{equation}
with
\begin{equation}\label{eq:ErrorN}
E_N=
\pm
\max_{0\le x\le x_{\rm max}}
\left|
\|x\|_2^2-\sum\limits_{j=1}^N\sigma_j u^2_j(x)\right|
\end{equation}
being positive (negative) if the left hand side of \eqref{eq:AlternantTheorem} is positive
(negative) at $x=x_0^*$.

The alternant theorem provides the basis for the non-linear Remez algorithm that
has been used successfully in other papers and yields the minimax solution
$\lbrace\sigma_j^*,\tau_j^*\rbrace_{j=1}^N$.\cite{Braess2005, Hackbusch2008,
Kaltak2014} The minimax solution also yields abscissas in the unscaled interval
$0\le\tau^*_j\le\frac12$ and the corresponding weights $\sigma^*_j$ are positive
and satisfy the sum rule $\lim_{N\to\infty}\sum_{j=1}^N \sigma^*_j =1$. This is
important for the application in many-body theory, since the conservation of
particles is guaranteed with increasing $N$ including particles with energy
$\epsilon_\alpha\approx\mu$. 

Before we discuss the construction of the frequency grids a last remark is in
place here. The quadrature obtained from the solution of
\eqref{eq:MinimaxProblemTime} is also a good approximation to the solution for
the corresponding problem for the odd basis function \eqref{eq:BasisTimeOdd}.
On the other hand, the linear combination of $u_\tau(x)$ and $v_\tau(x)$ yields
a similar basis $e^{-\frac{x}2(1-2|\tau|)}/\cosh\frac{x}2$ that has been used
recently by Shinaoka and coworkers to compress Green's functions on the
imaginary time axis in quantum Monte Carlo algorithms.\cite{PhysRevB.96.035147}
This, implies a close connection to our method. However, Shinaoka {\em et al.}
determine the grid as the solution of an integral equation and the connection to
$\varepsilon$-isometric subspaces is not immediately evident. 

\subsection{Bosonic Frequency Grid}\label{sec:BosonicFrequencyGrid}
To construct the bosonic frequency  grid we use the $\varepsilon$-isometric basis
of the even time basis \eqref{eq:BasisTimeEven}, specifically again the IA$_1$ basis 
\begin{equation}
\label{eq:BasisFreqEvenBos}
\overline u_{\nu_n}(x)=\frac{x}{x^2+\nu_n^2}\tanh\frac{x}2.
\end{equation}
The motivation behind this choice is three-fold. Firstly, it is obtained from
the cosine transformation of the even time basis \eqref{eq:BasisTimeEven}
and it is suitable for bosonic quantities. Thus it can describe the
imaginary frequency dependence of the polarizability \eqref{eq:IPPolarizability}
that is of bosonic nature; the IA$_2$ basis is obtained from the sine transform
of these functions and has fermionic symmetry and hence
irrelevant for the evaluation of bosonic integrals. Secondly, we can use the
minimax isometry method to switch between the frequency and time representation
of the polarizability with high precision.  This follows from the theorem proved
in section \ref{sec:MinimaxIsometry} Equ.  \eqref{eq:EpsilonIsometry}.  Lastly,
the infinite bosonic Matsubara series of the RPA grand potential
\eqref{eq:DefRPAGrandPotential} can be evaluated with high precision without
using any interpolation technique.

In practice, the unscaled bosonic frequency quadrature for $\beta=1$ is
determined first and following scaling relations are used to obtain the result
for arbitrary inverse temperatures
\begin{equation}\label{eq:ScalingFreq}
\nu_k \to \frac{\nu_k}\beta, \quad \lambda_k \to \frac{\lambda_k}\beta.
\end{equation}
The corresponding minimax problem reads
\begin{equation}\label{eq:MinimaxProblemA1}
\min_{\lambda_k>0,\nu_k\in(0,\infty)}
\max_{0\le x\le x_{\rm max}}
\left|
\|x\|_2^2- \sum\limits_{k=1}^{N}\lambda_k \overline u^2_{\nu_k}(x)
\right|
\end{equation}
where the $L^2$-norm $\|x\|_2^2$ is given in Tab. \ref{tab:IsometricBases}. The
solution $\lbrace \lambda_k^*,\nu_k^*\rbrace_{k=1}^N$ is called
IA$_1$-quadrature in the following, in agreement with the notation used in Tab.
\ref{tab:IsometricBases}.

\section{Frequency Grid for GM}\label{sec:FrequencyGM}

In this section, we discuss the construction of a compressed fermionic frequency
grid for selfconsistent Green's function calculations using  Equ.
(\ref{eq:DysonG}), and evaluation of the correlation energy using the GM
expression  for the grand potential \eqref{eq:DefGMGrandPotential}.

Ideally, quadratures of the GM expression should be  converging exponentially
with the number of grid points $N$.  In contrast to the polarization function
and second order correlation energies, this requires an accurate handling of
fermionic functions of the type IA$_2$ and IB$_1$ in the frequency domain, which
is an intricate problem.

Why the evaluation of  the GM energy is more difficult than calculation of the
correlation energy for the RPA is discussed in the appendix
\ref{sec:GMdifficult} in detail. The problem is, however, also  obvious, when
we desire to calculate the self-consistent Green's function from the self-energy
using the Dyson equation [compare Equ. \eqref{eq:DysonG}]. The optimization of
the frequency grid yields widely different frequencies for the symmetric and
anti-symmetric part of the Green's function.  However clearly, in order to solve
the Dyson equation, we need both the symmetric and anti-symmetric part of the
Green's function on the same frequency grid. Attempts to chose one grid over the
other yields slow convergence of the total correlation energy.  A solution to
this dilemma is presented in the following section. 

\subsection{Fermionic Frequency Grid via $L^1$-norm}\label{sec:FermionicFrequencyGrid}

Every fermionic Matsubara series of a function $\tilde A$ (e.g. $\tilde A=\tilde G(z) \tilde \Sigma(z)$),
$\sum_{n\in\BoldMath{Z}} \tilde A(\omega_n)$, corresponds to a complex contour
integral of that function times the Fermi function (see derivation below or
Fetter and Walecka\cite{Fetter2003}).  Hence, finding approximations of the
Fermi function with as few poles as possible accelerates the calculation of any
Matsubara series by replacing the Matsubara summation by a summation
over the poles of the approximated Fermi function.

This idea was exploited by Ozaki\cite{Ozaki2007} in combination with the
following identity for the Fermi function
\begin{equation}\label{eq:FermiTanh}
  f(x)-\frac{1}{2} =    \frac12\tanh\frac{x}{2} = \sum\limits_{m\in\BoldMath{Z}}
  \underbrace{\frac{x}{x^2+\omega_m^2}}_{\tilde u_{\omega_m}(x)}.
\end{equation}
A proof of this identity is found in the appendix \ref{app:ProofTanh}.  Ozaki
used a  partial fraction decomposition of the hyperbolic tangent in combination
with a continued fraction representation of the hypergeometric function $_1F_0$
to derive a compressed form of \eqref{eq:FermiTanh}.

Our approach is based on the observation that the  $L^1$-norm of our previously
defined basis functions  $\tilde u_{\omega_m}(x)$
\begin{equation}\label{eq:L1NormDef}
\|x\|_1=\sum\limits_{m\in\BoldMath{Z}}|\tilde u_{\omega_m}(x)| =\frac12\tanh\frac{|x|}{2},
\end{equation}
is also equivalent to the hyperbolic tangent and thus the Fermi function.  This suggests to
determine the frequency points and weights by solving the following minimization
problem:
\begin{equation}\label{eq:MinimaxProblemF}
\min_{\gamma_k,\omega_k>0}
\max_{0\le x\le x_{\rm max}}
\left|
\|x\|_1-\sum\limits_{k=1}^{N}\gamma_k |\tilde u_{\omega_k}(x)|
 \right|.
\end{equation}
Clearly this is very similar to Equ. (\ref{eq:MinimaxProblemA1}), but replaces
the $L^2$- by the $L^1$-norm.  Because $\|x\|_2\le\|x\|_1$ holds true for any
$0\le x\le x_{\rm max}$,\cite{Fleming2002} the $L^1$-solution
$\lbrace\gamma_k^*,\omega_k^*\rbrace_{k=1}^N$, called F-quadrature in the
following,  yields linearly independent basis functions $\tilde u^*_k$ that span
a larger function space than the basis obtained from
corresponding $L^2-$solutions discussed in the appendix \ref{sec:GMdifficult}.
Our numerical experiments presented below show that the F-quadrature evaluates
the infinite sum over both, even and odd functions [see Equ.
\eqref{eq:GMMatrixElement}] with high precision for increasing $N$. 

Both, the F-quadrature and Ozaki's hypergeometric quadrature (OHQ), use
essentially a rational polynomial approximation to the hyperbolic tangent. In
the following, we show why this approach also provides a good approximation of
fermionic Matsubara series, such as the the density matrix $\Gamma$ for holes (upper
sign) and electrons (lower sign). The density matrix $\Gamma$ satisfies the following identity
\begin{equation}\label{eq:DensityMatrix}
\Gamma =\pm\lim_{\eta\to0\pm} G(-i\eta)=\pm\lim_{\eta\to0\pm}\frac1\beta\sum\limits_{n=-\infty}^{\infty}
\tilde G(i\omega_n)e^{-i\omega_n\eta},
\end{equation}
where $G$ and $\tilde G$ is the interacting Green's function in imaginary time
and on the Matsubara axis, respectively. Specifically, we show that the last
expression on the right hand side of \eqref{eq:DensityMatrix} can be approximated with
the following quadrature formula
\begin{equation}\label{eq:DensityMatrixQuadrature}
\Gamma \approx \frac{\mathrm{sgn}(\eta)}{2}\unit + \sum\limits_{k=1}^N
\frac{\sigma_k}2\left[\tilde G(i\omega_k)+\tilde G(-i\omega_k)\right],
\end{equation}
where $\unit$ is the identity matrix in the considered basis
and $\sigma_k,\omega_k$ are either the OHQ- or F-quadrature points. 
\begin{figure}[tbp]
\centering
\MyFigure{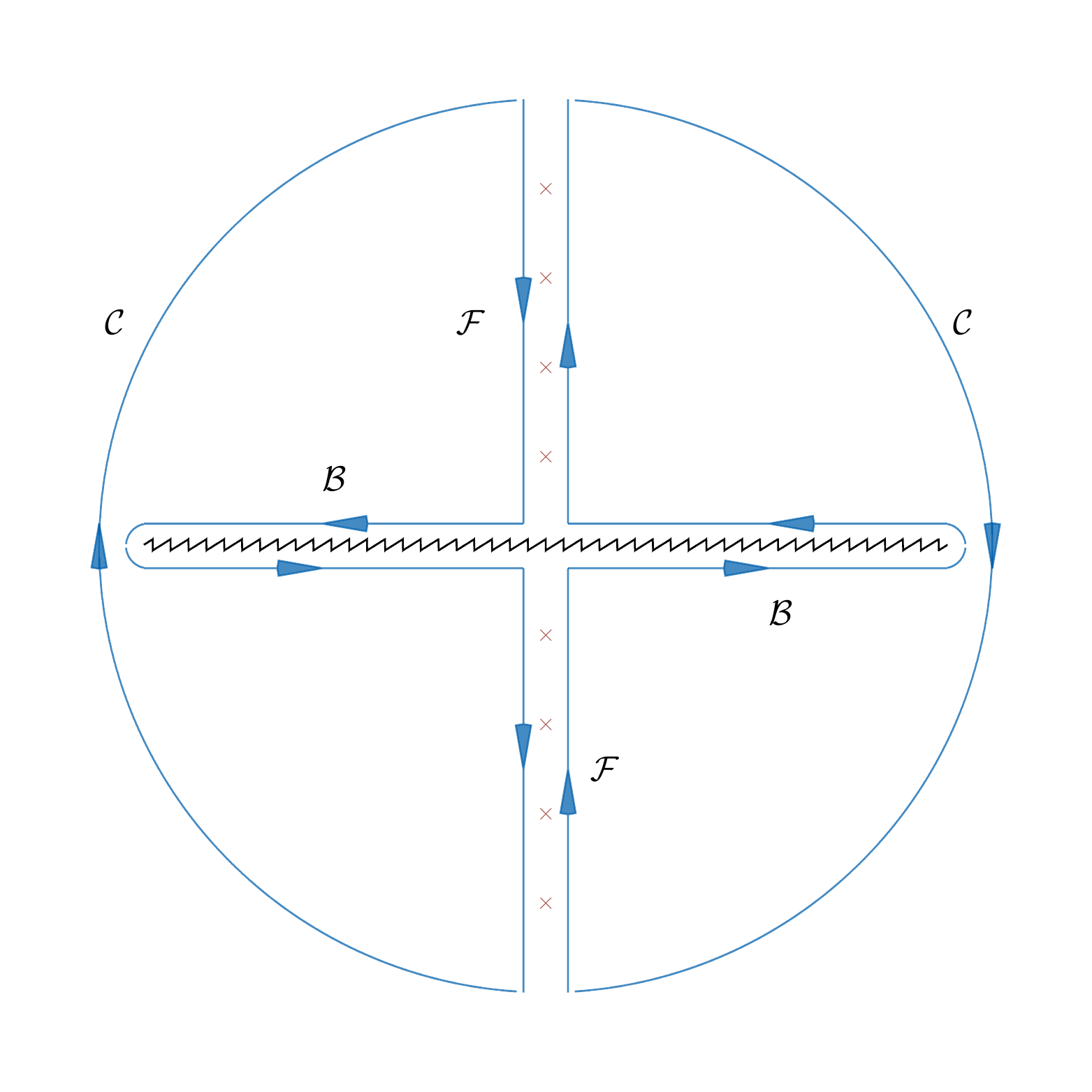}
\caption{ Integration contours in
\eqref{eq:ComplexContourIdentities}
(zigzag line) branch cut of $\tilde A$, (crosses) fermionic Matsubara
frequencies $\omega_n$ correspond to poles $z=i\omega_n$ of auxiliary function
$h_\eta(z)$ defined in \eqref{eq:AuxiliaryTanh} and \eqref{eq:AuxiliaryForG}
such that contour integral $\oint\mathrm{d}z \tilde A(z) h_\eta(z)$ for path
$\mathcal{C}$ is zero.
}
\label{fig:Contour}
\end{figure}

To derive  Equ. \eqref{eq:DensityMatrixQuadrature} and motivate why an
approximation to the hyperbolic tangent provides an excellent approach to
compress any fermionic Matsubara series, we consider a general correlation
function $\tilde A$ that is analytic in the complex plane $z$ with a branch cut
on the real axis and decays with $\mathcal{O}(|z|^{-1})$ or faster to zero for
$|z|\to\infty$. As examples, we consider  $\tilde G(z)$ and $\tilde G(z) \tilde
\Sigma(z)$.  Following Fetter and Walecka,\cite{Fetter2003} one introduces an
auxiliary function $h_\eta(z)$ with an infinitesimal $\eta$ to force the complex
contour integral over the infinite large outer circle $\mathcal{C}$ in
\fref{fig:Contour} to vanish, that is
\begin{equation}\label{eq:ComplexContourC}
\oint_{\mathcal{C}}\frac{\mathrm{d}z}{2\pi i}\tilde A(z)h_\eta(z) =0.
\end{equation}
Regardless of the specific choice of $h_\eta(z)$ (discussed below), one can 
easily show using the residue theorem and the contours depicted in
\fref{fig:Contour} the following identity:
\begin{equation}\label{eq:ComplexContourIdentities}
\begin{split}
\sum\limits_{n\in\BoldMath{Z}}\Res_{z=i\omega_n}
\left[\tilde A(z)h_\eta(z)\right]
=&\oint_{\mathcal{F}}\frac{\mathrm{d}z}{2\pi i}\tilde A(z)h_\eta(z)\\
=&-\oint_{\mathcal{B}}\frac{\mathrm{d}z}{2\pi i}\tilde A(z)h_\eta(z)\\
=&\int_{-\infty}^\infty \mathrm{d}\omega\frac1\pi\mathrm{Im}\left[\tilde
A(\omega)h_\eta(\omega)\right]
\end{split}
\end{equation}
Apart from condition \eqref{eq:ComplexContourC}, the auxiliary function
$h_\eta(z)$ has to be chosen such that the left hand side in
\eqref{eq:ComplexContourIdentities} gives the fermionic Matsubara series
$\sum_{n\in\BoldMath{Z}}\tilde A(i\omega_n)e^{-i\omega_n\eta}$, which imposes two
conditions on $h_\eta$.  Firstly, $h_\eta$ must have an infinite number of poles
located at $z=i\omega_n$ (crosses in \fref{fig:Contour}). Secondly, the
corresponding residues have to be $\pm \tilde A(i\omega_n)e^{-i\omega_n\eta}$ for
$\eta\to0\pm$. 
 
If $\tilde A(z)$ is of order $\mathcal{O}(|z|^{-1-\delta}),\delta>0$ for
$|z|\to\infty$ (e.g. $\tilde A(z)=\tilde G(z)\tilde \Sigma(z)$) the outer
contour integral \eqref{eq:ComplexContourC} is zero, even for the simplest
choice for $h_\eta(z)$, specifically,
\begin{equation}\label{eq:AuxiliaryTanh}
\begin{split}
h_\eta(z)=&\frac12\tanh\frac{z}2\\
=&\frac12\sum\limits_{n\in\BoldMath{Z}}\left[\frac1{z-i\omega_n}+\frac1{z+i\omega_n}\right],
\end{split}
\end{equation}
where the last line follows from \eqref{eq:FermiTanh} and reflects the locations
and residue of the poles of $h_\eta(z)$.  Approximating the hyperbolic tangent
by a rational polynomial with poles on the imaginary axis allows one to find
accurate approximations for the right hand side in
(\ref{eq:ComplexContourIdentities}), by replacing the Matsubara series  (left
hand side in Equ. (\ref{eq:ComplexContourIdentities})) by a  sum over the poles
of the rational approximation of the $\tanh(z/2)$.

In contrast, for correlation functions $\tilde A$ that decay only with
$\mathcal{O}(|z|^{-1})$ at $|z|\to\infty$ the sign of the infinitesimal $\eta$
matters. This includes $\tilde G(z)$ as well as any mean field terms, e.g.
Green's function times the mean field Hamiltonian $\tilde G(z) H_0$. For
instance, the limit $\eta\to0-$ in \eqref{eq:DensityMatrix} gives the electron
density matrix, while $\eta\to0+$ gives the density matrix of holes. For
functions of order $\mathcal{O}(|z|^{-1})$ at $|z|\to\infty$, one therefore has
to add a term to the hyperbolic tangent. As can be shown easily, the form for
$h_\eta(z)$ for which \eqref{eq:ComplexContourC} holds true
is\cite{Fetter2003}
\begin{equation}\label{eq:AuxiliaryForG}
h_\eta(z)=\left[\frac{\mathrm{sgn(\eta)}}{2}+\frac12\tanh\frac{z}2\right]e^{-z\eta}.
\end{equation}
Inserting Equ. \eqref{eq:AuxiliaryForG} into the right hand side of
\eqref{eq:ComplexContourIdentities} yields
\begin{equation}\label{eq:ComplexContourDensity}
\begin{split}
\sum\limits_{n\in\BoldMath{Z}}
\tilde A(i\omega_n)e^{-i\omega_n\eta}=&
\frac{\mathrm{sgn}(\eta)}2
\int_{-\infty}^\infty
\mathrm{d}\omega\frac1\pi\mathrm{Im}\left[\tilde A(z)e^{-z\eta}\right] \\
+&
\int_{-\infty}^\infty
\mathrm{d}\omega\frac1\pi\mathrm{Im}\left[\tilde
A(z)\frac12\tanh\frac{z}2e^{-z\eta}\right],
\end{split}
\end{equation}
In the last term on the right hand side the evaluation of the limit $\eta\to 0$
can be performed before integration, because the integrand is of order
$\mathcal{O}(|z|^{-2})$ for $|z|\to\infty$  [see Equ. \eqref{eq:AuxiliaryTanh}].
The corresponding integral over the arch $\mathcal{C}$ vanishes, so that the
last term in \eqref{eq:ComplexContourDensity} on the right hand side  can be
rewritten into the Matsubara series of $\tilde A$ that is independent of the
sign of $\eta$.  This is the {\em convergent part} of the Matsubara series and
the term that can be again evaluated using the rational approximation of the
$\tanh$ and quadratures. In contrast, the first term on the right hand side of
\eqref{eq:ComplexContourDensity} cannot be written into a Matsubara series,
because the integrand diverges for $|z|\to\infty$ prohibiting the closure of the
integration contour at infinity.  However, for $\tilde A(z) = \tilde G(z)$ one
has 
\begin{equation}\label{eq:DivergentTermExplicit}
\lim_{\eta\to0\pm}\frac{\mathrm{sgn}(\eta)}{2}
\int_{-\infty}^\infty
\mathrm{d}\omega\frac1\pi\mathrm{Im}\left[\tilde G(z)e^{-z\eta}\right] =\pm
\frac{1}{2}\unit,
\end{equation}
which concludes our proof of Equ. \eqref{eq:DensityMatrixQuadrature}. 
We call this term, therefore, the {\em divergent part} of the Matsubara series, 
although the term is finite in any practical calculation (number of
electrons/holes is finite in practice). 
We use \eqref{eq:DensityMatrixQuadrature} for the evaluation of the density
matrix in self-consistent $GW$ calculations at finite temperature (see section
\ref{sec:Results}).

In summary, the approximation of the hyperbolic tangent by rational polynomials
with poles only on the imaginary axis gives rise to fermionic frequency
quadratures that describe the convergent part of the Matsubara series. To obtain
the directional limits $\eta\to0\pm$ of slowly decaying correlation functions,
such as the propagator of electrons or holes, the integral over the spectral
function of the integrand has to be added or subtracted, respectively. The
evaluation of the GM energy does not require this term, because $\tilde
G(z)\tilde \Sigma(z)$ decays with $\mathcal{O}(|z|^{-2})$. Analogous bosonic
quadratures can be obtained by approximating the $L^1$-norm of the hyperbolic
cotangent, but this was not further investigated. 
\begin{figure}[tbp]
\centering
\MyFigure{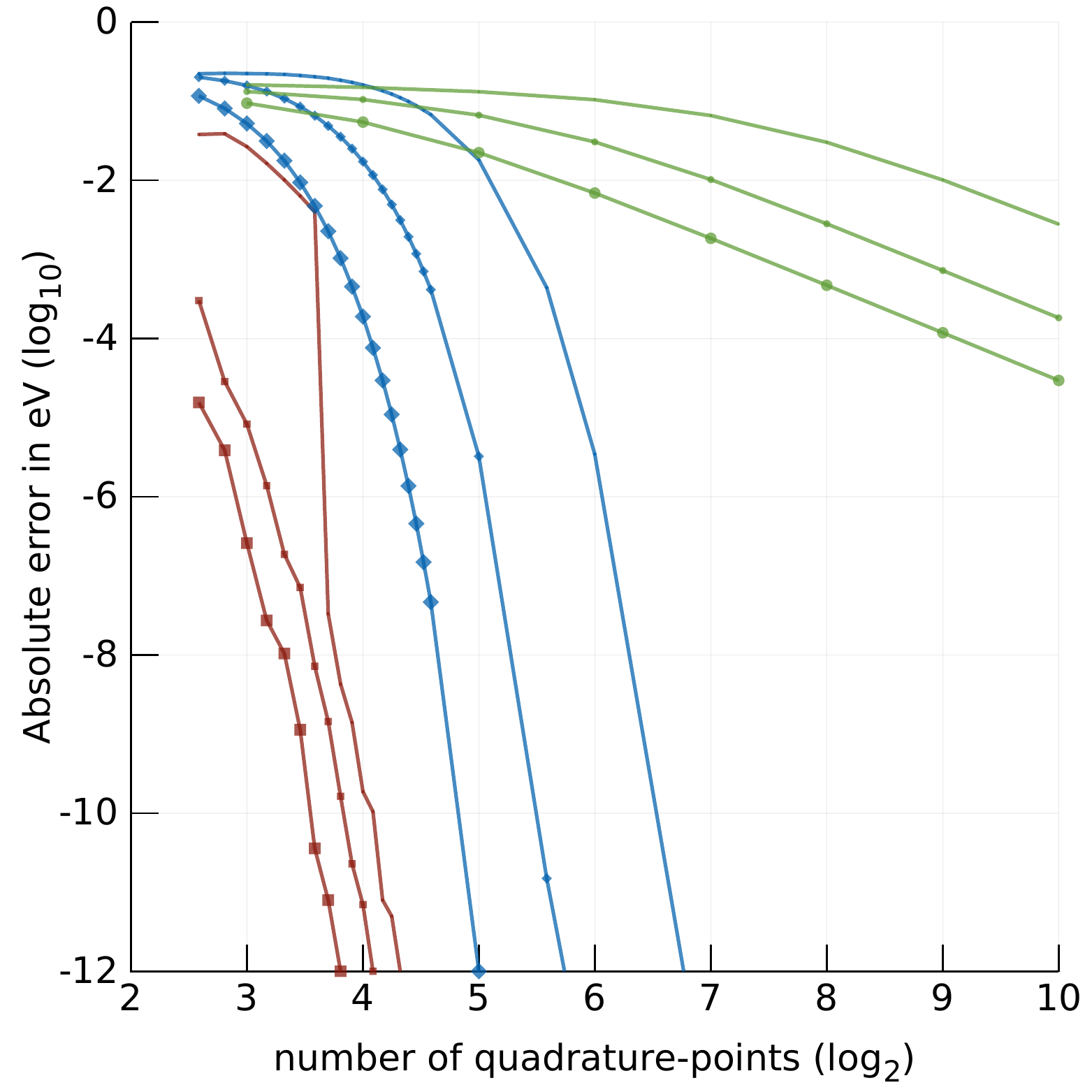}
\caption{Convergence of Matsubara grid (points), hypergeometric grid (diamonds)
and F-grid (squares) for inverse temperatures $\beta=1,10,100$
eV$^{-1}$ (small, medium, large symbols).  
}
\label{fig:MatsubaraConvergence}
\end{figure}

We have compared our F-quadrature with Ozaki's hypergeometric quadrature (OHQ) by
means of calculating the GM factor \eqref{eq:GMMatrixElement} for
a model that includes 50 randomly sampled poles in $-0.05\le x,y\le 0.05$ and 50
poles in $-50\le x,y\le 50$. The results for $\beta=1,10$ and $100$ eV$^{-1}$
are shown in \fref{fig:MatsubaraConvergence} and are contrasted to the grid
convergence for the ordinary fermionic Matsubara quadrature $\lbrace
\gamma_m=2,\omega_m=(2m-1)\pi/\beta\rbrace_{m=1}^N$. 
It can be seen that the F-quadrature outperforms the OHQ in all cases,
especially for $\beta>1$ (low temperatures). This can be explained by the fact
that the F-grid minimizes the quadrature error for all energies uniformly in the
interval $|x|,|y|\le x_{\max}$. The corresponding OHQ-quadrature error is
non-uniformly distributed in the same interval and has the effect that at high
$\beta$ values the convergence is very slow with the number of grid points
for small $N$. The same figure, also shows the linear convergence of the
conventional Matsubara grid and demonstrates its pathology in practice. 

\section{Discrete Time to Frequency Transformations}\label{sec:MinimaxFourierTransformation}
\subsection{Minimax Isometry Transformation}\label{sec:MinimaxFourierTransformation}
We have seen how different basis functions for the time and frequency domain
give rise to different grids. In this section we study the error made by
transforming an object represented on the time grid
$\lbrace\tau_1^*,\cdots,\tau^*_N\rbrace$ to the frequency
axis. As a measure for the transformation error we use
\begin{equation}\label{eq:TransformationErrorPhi}
\tilde E(\omega)=
\min_{t_{\omega k\in\BoldMath{R}}}
\left\|\mel*\omega{x}-\sum_{k=1}^Nt_{\omega k}\mel*{\tau_k^*}x 
\right\|_2^2,
\end{equation}
where $\mel*{\tau^*_k}x$ is here always the even time basis function
\eqref{eq:BasisTimeEven} (IA$_1$ in Tab. \ref{tab:IsometricBases}) evaluated at
the minimax time grid obtained from \eqref{eq:MinimaxProblemTime} and
$\mel*{\omega}x$ acts as a placeholder for one of the basis functions in the
frequency domain listed in Tab. \ref{tab:IsometricBases} with $\omega$ being a
positive real number.  The $L^2$-norm is evaluated by sampling the $x-$values
with 100 points $\xi_j^*$ determined from the alternant $\lbrace
x_j^*\rbrace_{j=0}^{2N}$ of the minimax time problem in
\eqref{eq:MinimaxProblemTime} and additional $(101-2N)/2N$ uniformly distributed
points in each of the $2N$ sub-intervals $[x_j^*,x_{j+1}^*]$. 

The solution of the ordinary least square problem for the frequency $\omega$ is
then given by the corresponding normal equation\cite{NR2007}
\begin{equation}\label{eq:TransformationErrorSolution}
\begin{split}
&\sum_{i=1}^{100}\mel*\omega{\xi_i^*}\mel*{\xi_i^*}{\tau_k^*} = \\
&\sum_{j=1}^{N}t_{\omega j}\sum_{i=0}^{100}\mel*{\tau_j^*}{\xi_i^*}\mel*{\xi_i^*}{\tau_k^*},\quad 
k=1,\cdots,N.
\end{split}
\end{equation}
Solving this equation yields the desired discrete time-to-frequency
transformation coefficients $t_{\omega j}$.  The transformation error is rather
insensitive to changes of the number of sampling points $\xi_j^*$; the $2N+1$
alternant points $x_j^*$ of the time grid also often suffice in practice.

\begin{figure}[tbp]
\centering
\MyFigure{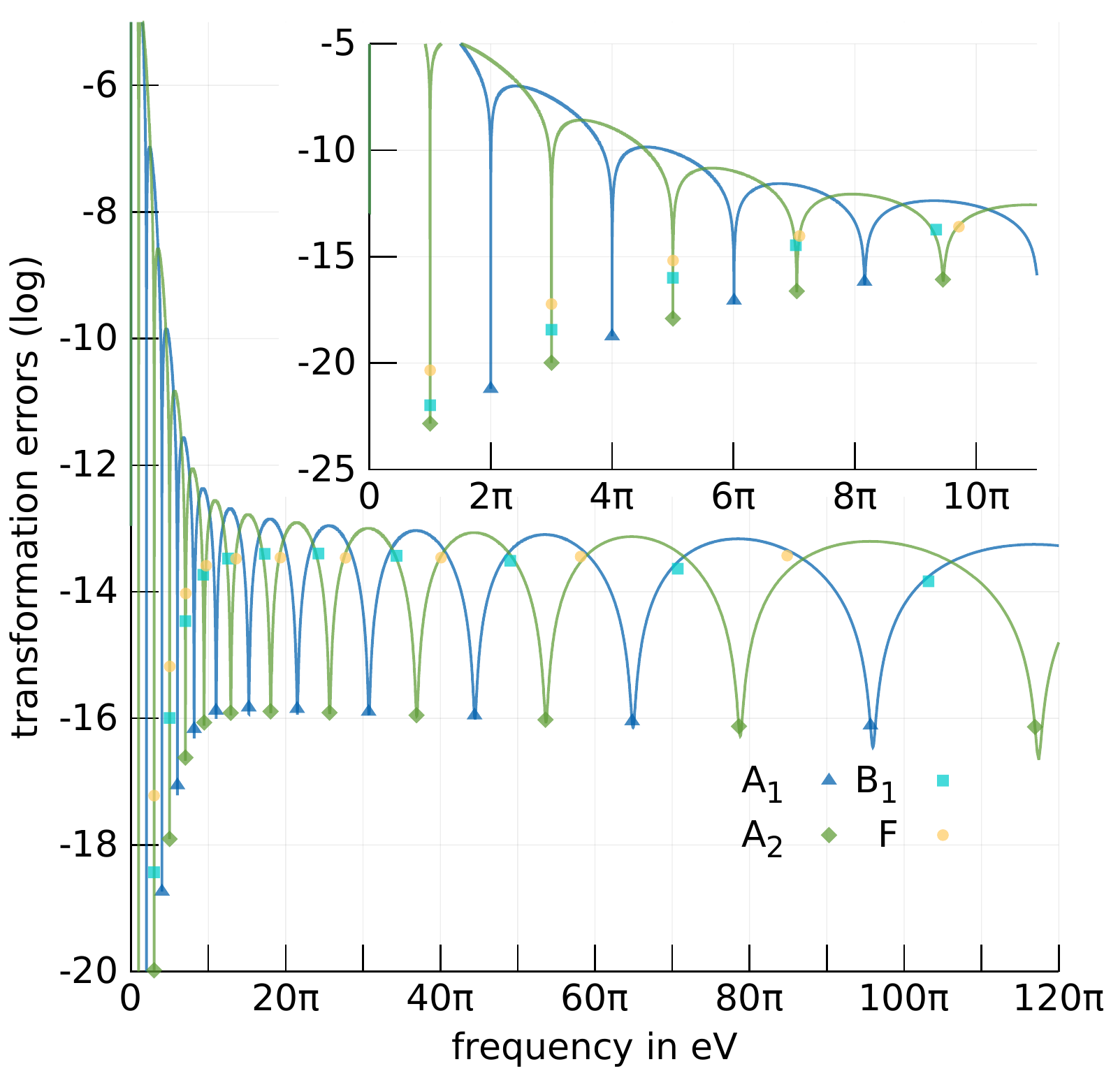}
\caption{ Transformation error $\tilde E(\omega)$ from even time basis functions
\eqref{eq:BasisTimeEven} to the IA$_1$ (blue) and IA$_2$ (green) frequency basis
for $N=16$, $x_{\max}=1000$.  The inset shows the low frequency regime.  Points
indicate minimax grid points in frequency domain for IA$_1$,  IA$_2$, IB$_1$ and
F (the abscissa corresponds to the optimal frequency, whereas  the ordinate is
given by the error $\tilde E(\omega)$ determined at the respective frequency point).}
\label{fig:TransformationErrors}
\end{figure}

In the normal modus operandi, we would solve for $t_{\omega j}$ at a set of
previously chosen frequencies $\omega$.  However, Equ.
\eqref{eq:TransformationErrorPhi} also allows to plot $\tilde E(\omega)$ as a
function of the frequency $\omega$. This gives independent insight, on which
frequencies one is supposed to use in combination with a certain set of time
basis functions, independent of the previous considerations (see
\fref{fig:TransformationErrors}).

Transformation to the IA$_1$ frequency basis functions
\eqref{eq:SymmetricPartChiInFreq} (blue line), clearly
shows that the error $\tilde E(\omega)$ is minimal at the previously determined
IA$_1$ frequency points (blue triangles), and transformation to the IA$_2$
frequency basis functions (green line) shows that the error is smallest at the
previously determined IA$_2$ frequency points (green diamonds).  The reason for
this behavior is due to the fact that the IA$_1$, IA$_2$ and the time quadrature
for the even time basis  $ u_\tau(x)$ [Equ. \eqref{eq:BasisTimeEven}] possess
the same approximation property and span $N$-dimensional, almost isometric
subspaces of the Hilbert space $\mathcal{H}$ as proven in section
\ref{sec:MinimaxIsometry}.  The good agreement is a numerical confirmation that
the previously determined frequency grids are optimal.

Thus, for polarizabilities and second order correlation energies, the optimal
frequency points are clearly the IA$_1$-quadrature points (triangles). They
approach the conventional bosonic frequencies $2m \pi$ [Equ.
\eqref{eq:BosonicFrequencies}] at small $\omega$. The corresponding weights (not
shown) approach $2$ (except for the first frequency $\nu^*_1=0,~\lambda^*_1=1$).
Higher quadrature frequency points (as well as weights) deviate considerably
from the conventional bosonic Matsubara points $ \nu_m = 2\pi m$. This behavior
is very similar to the bosonic grid presented by Hu {\em et al.} that is based
on the continued fraction decomposition of the hyperbolic cotangent, the
analogue of Ozaki's method for bosons.\cite{Hu2010} However, the
IA$_1$-quadrature has the advantage that the error is minimized uniformly for
all transition energies $|x|\le \beta \epsilon_{\max}$, while the continued
fraction method yields non-uniformly distributed errors in general. 

Transformation from the even time basis  to the fermionic frequency basis IA$_2$
[Equ.\eqref{eq:AntisymmetricPartG0InFreq}] yields further insight. As already
emphasized, the minimax IA$_2$ frequency points match exactly those frequency
points where the error for transformation into the IA$_2$ basis functions is
minimal. On the other hand, the  IB$_1$ minimax grid points are chosen to
optimally represent odd time basis functions $v_\tau(x)$ [Equ.
\eqref{eq:BasisTimeOdd}] using the corresponding frequency basis [Equ.
\eqref{eq:SymmetricPartG0InFreq}].  At small frequencies, these points are
slightly shifted away from the optimal IA$_2$ frequency points, resulting in
somewhat larger transformation errors. This is to be expected, since the points
have been chosen to approximate a different scalar product than for IA$_2$ (and
IA$_1$), see fourth column in Tab.  \ref{tab:IsometricBases}. Specifically, the
IB$_1$ minimax frequencies are by construction optimal to represent odd time
basis functions.  Although,  IB$_1$ and IA$_2$ minimax points are close at low
frequencies, they progressively move away at higher frequencies, which prohibits
the construction of a common frequency grid for fermions.

From figure \ref{fig:MatsubaraConvergence}, it is somewhat unclear why
the F frequency grid works well, although it is noteworthy that the
corresponding frequency points lie roughly at the positions where IA$_1$ and
IA$_2$ errors intersect. This might imply an equally acceptable representation
of odd and even fermionic functions at the cost of larger errors. 

\subsection{$\varepsilon$-isometric time grids of the
F-quadrature}\label{sec:EpsilonIsometricTimeGrid}
\begin{figure}[tbp]
\centering
\MyFigure{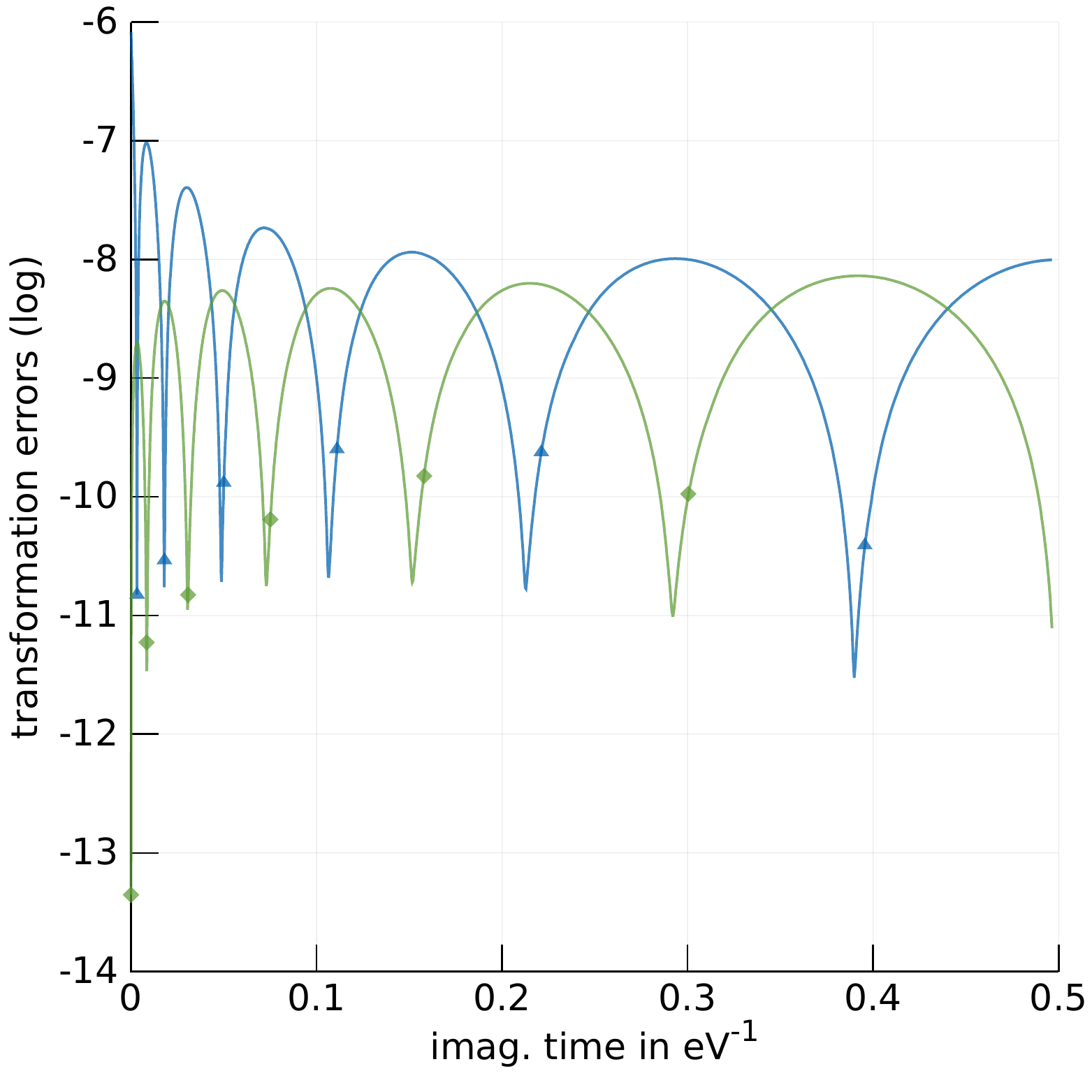}
\caption{ Transformation error $E(\tau)$ from frequency F-grid to time domain
for the IA$_2$ (blue line) and IB$_1$ (green line) $\varepsilon$-isometric basis
functions in Tab. \ref{tab:IsometricBases} for $N=6$, $x_{\max}=100$.
Triangles and diamonds indicate the IA and  IB time grids, respectively.
}
\label{fig:TransformationErrorsF}
\end{figure}

Recapitulating the previous section, a natural question arises: Is there an
optimum time grid for the F-quadrature?  In analogy, to Equ.
\eqref{eq:TransformationErrorPhi} this grid may be defined by the minima of the
inverse transformation error
\begin{equation}\label{eq:InverseTransformationError}
E(\tau)=
\min_{t_{\tau k\in\BoldMath{R}}}
\left\|\mel*\tau{x}-\sum_{k=1}^Nt_{\tau k}\mel*{\omega_k^*}x 
\right\|_2^2,
\end{equation}
where $\omega_k^*$ are the abscissa of the F-quadrature. Table
\ref{tab:IsometricBases} shows that there are two possible choices for the 
transformation 
$\mel*{\omega_k^*}{x}\to \mel*{\tau}{x}$; one
function describing the transformation error for the IA$_2$  and one for IB$_1$
basis functions given in Tab. \ref{tab:IsometricBases}. 
Both transformation errors
are plotted in \fref{fig:TransformationErrorsF} (blue and green line,
respectively). 
 
The figure clearly shows that the minima of both error functions differ and
implies two $\varepsilon$-isometric time grids for the frequency F-grid. This is
analogous to the forward transformation errors discussed in the previous section,
where the IA$_1$ and IA$_2$ frequency grids are made up by widely different
frequencies. For the F-grid, however, the IA$_2$ and IB$_1$ minimax solutions in
time coincide with the minima of the error functions only for $\tau\approx0$,
for larger values of $\tau$ the transformation error minima
($\varepsilon$-isometric grids) deviate from the corresponding minimax grid
points (compare minima of green and blue curve with triangles and diamonds in
\fref{fig:TransformationErrorsF}).

The small IB$_1$ transformation error for $\tau=0$ follows from the fact that for
small $\tau$ the time basis becomes $\mel*{\tau}x=v_\tau(x)\approx\tanh (x/2)$.
Per construction [see Equ. \eqref{eq:L1NormDef}], the hyperbolic tangent function is
approximated well by the basis \eqref{eq:SymmetricPartG0InFreq} using the F-grid.
The IA$_2$ transformation error (blue line), in contrast, is several orders of
magnitude larger at $\tau=0$, since the time basis function is constant
$\mel*{\tau=0}x= u_{\tau=0}(x)=1$ and the frequency basis functions
$\mel*{\omega_k}x$ represent constants only poorly.  The deviation of the IA$_2$ and
IB$_1$ time grids from the $\varepsilon$-isometric time grids at higher $\tau$
values is not surprising, since the F-grid deviates from the
$\varepsilon$-isometric frequency grids, that is the IA$_2$ and IB$_1$-grid
discussed in section \ref{sec:BosonicFrequencyGrid} and
\ref{sec:FermionicFrequencyGrid}. 

In summary, we recommend using the IA$_1$ time grid presented in section
\ref{sec:TimeGrid} in combination with the IA$_1$ frequency grid for bosonic
functions (see \ref{sec:BosonicFrequencyGrid}) and the F-grid for fermionic
functions. First, the exact $\varepsilon$-isometric time points of the F-grid are
only known numerically from inspection of the transformation error; an analogue
to the minimax isometry method is not known to us.  Second, Green's functions in
imaginary time can be contracted without error, whilst at the same time,
transformation errors to the imaginary frequency axis are controlled. Before we
demonstrate these advantages in section \ref{sec:Results}, we discuss the
following details about our implementation in the Vienna ab initio software
package (VASP).\cite{PhysRevB.59.1758}

\section{Technical Details}\label{sec:TechnicalDetails}
The implementation of the finite temperature RPA and $GW$ algorithms is the same
as the zero temperature ones,\cite{Kaltak2015,PhysRevB.94.165109} with three
exceptions. 
\begin{itemize}
\item The zero temperature frequency grid is replaced by the IA$_1$-grid
discussed in \ref{sec:BosonicFrequencyGrid} for the bosonic correlation
functions $\tilde\chi,\tilde{W}$, while the F-quadrature from
\ref{sec:FermionicFrequencyGrid} replaces the grid for the fermionic functions
$\tilde{G}$ and $\tilde\Sigma$. 
\item All correlation functions are evaluated on the same imaginary time grid
presented in sec.  \ref{sec:TimeGrid}.
\item The occupied and unoccupied Green's function $\underline{G},\overline{G}$
need to be set up carefully considering the partial occupancies $f(x_\alpha)$ in each
system.
\end{itemize}
The last point requires some clarification. The Green's function for positive
$\overline G$ and negative times $\underline{G}$ can be combined to a full
Green's function using Heaviside theta functions
\begin{equation}\label{eq:DefGreensFunction}
G(-i\tau) = \Theta(\tau)\overline{G}(\tau)-\Theta(-\tau)\underline{G}(\tau).
\end{equation}
At zero temperature $(\beta\to\infty)$ the occupied and unoccupied imaginary
time Green's function read\cite{Rojas1995,Kaltak2015}
\begin{align}\label{eq:DefGoccT0}
\underline{G}(\tau)|_{\beta=\infty}
=&\sum\limits_{\alpha}\Theta(-x_\alpha)e^{-x_\alpha\tau}
\\
\overline{G}(\tau)|_{\beta=\infty}
=&\sum\limits_{\alpha}\Theta(+x_\alpha)e^{-x_\alpha \tau}
\label{eq:DefGunoT0}.
\end{align}
Here the step function $\Theta$ ascertains that $\overline{G}$ and $\underline{G}$
contains only unoccupied (occupied) one-electron states. This changes as
temperature increases, because the step function $\Theta$ is replaced by the
Fermi function
\begin{equation}\label{eq:HeavisideTheta}
\Theta(\pm x_\alpha)\to
f(\mp x_\alpha)=\frac{1}{e^{\mp \beta x_\alpha}+1},
\end{equation}
implying the form given in \eqref{eq:DefG0} for the full Green's function
\eqref{eq:DefGreensFunction}. Consequently, the Green's function $\underline G$
needs to include also partially occupied states at finite temperature and vise
versa, so that the positive and negative imaginary time Green's functions
\begin{align}\label{eq:DefGneg}
\underline{G}(\tau)
=&\sum\limits_{\alpha}f(+x_\alpha) e^{-x_\alpha\tau},\quad\tau<0\\
\overline{G}(\tau)
=&\sum\limits_{\alpha}f(-x_\alpha) e^{-x_\alpha\tau},\quad\tau>0
\label{eq:DefGpos}
\end{align}
are determined instead and include all considered one-electron states. 

We emphasize that for $-\beta\le \tau\le \beta$ there are no exponentially
growing terms, in neither of the two Green's functions, because of the simple
property of the Fermi function
\begin{equation}\label{eq:FermiFunctionReversal}
f(x_\alpha) = [1-f(x_\alpha)]e^{-x_\alpha\beta}.
\end{equation}
In agreement with the Feynman-St\"uckelberg interpretation of
QFT,\cite{RevModPhys.20.367,Stueckelberg1941} every
occupied state ($x_\alpha<0$) in the positive time Green's function
$\overline{G}$ is essentially a state propagating negatively in time 
\begin{equation}\label{eq:TimeReversalUnoccupiedStates}
[1-f(x_\alpha)]e^{-x_\alpha\tau} =
f(x_\alpha) e^{-x_\alpha(\tau-\beta)}, \quad 0<\tau<\beta
\end{equation}
and vise versa for $x_\alpha>0$ and negative times
\begin{equation}\label{eq:TimeReversalOccupiedStates}
f(x_\alpha) e^{-x_\alpha\tau} =
[1-f(x_\alpha)]e^{-x_\alpha(\tau+\beta)}, \quad -\beta<\tau<0.
\end{equation}
Note, that all time points of the constructed time grid in section
\eqref{sec:TimeGrid} obey $0<\tau^*_j<\frac\beta2$, such that the restrictions for
$\tau=\pm\tau^*_j$ in \eqref{eq:TimeReversalUnoccupiedStates} and
\eqref{eq:TimeReversalOccupiedStates} are never violated, respectively. 

\subsection{Computational details}\label{sec:Details}
The results presented in the following section have been obtained with VASP
using a $\Gamma$-centered k-point grid of $4\times4\times4$ sampling points in
the first Brillouin zone. To be consistent with the QFT formulation, Fermi
occupancy functions are forced by the code for all finite temperature many-body
algorithms (selected with \texttt{LFINITE\_TEMPERATURE=.TRUE.}), that is
\texttt{ISMEAR=-1} and the temperature in eV is set via the k-point smearing
parameter \texttt{SIGMA}. All calculations have been performed at  experimental
lattice constants of $a=5.431$ \AA~ for Si\cite{Levinstein1999} and $a=3.842$
\AA~ for SrVO$_3$\cite{Onoda1991}, respectively. For both, Si as well as
SrVO$_3$ the non-normconserving $GW$ potentials released with version 5.4.4,
specifically \texttt{Si\_sv\_GW, Sr\_sv\_GW, V\_sv\_GW} and \texttt{O\_s\_GW}
have been used and energy cutoffs of 475.1 eV and 434.4 eV for the basis set
have been employed, respectively. This allows us to study the grid convergence in the
presence of semi-core states and yields results that can be extrapolated to
normconserving potentials with higher cutoffs.\cite{PhysRevB.90.075125} The
independent electron basis required for RPA and $GW$ calculations has been
determined with density functional theory in combination with the
Perdew-Burke-Ernzerhof functional\cite{Perdew.105.9982} and the $q\to0$
convergence corrections\cite{PhysRevB.73.045112} have been neglected.
Furthermore, because the polarizability converges faster with the number of
plane waves considered compared to the wavefunction,\cite{PhysRevB.77.045136}
smaller energy cutoffs of 316.6 eV and 289.6 eV for $\chi$ (set using
\texttt{ENCUTGW}) for Si and SrVO$_3$ have been chosen, respectively.

\section{Results}\label{sec:Results}
\subsection{ Performance of IA$_1$-quadrature for RPA } 
We have used the IA$_1$-quadrature to generalize our cubic scaling RPA
algorithm\cite{Kaltak2015} to finite temperatures in order to calculate the RPA
grand potential for SrVO$_3$ and Si. 

\begin{figure}[tbp]
\centering
\MyFigure{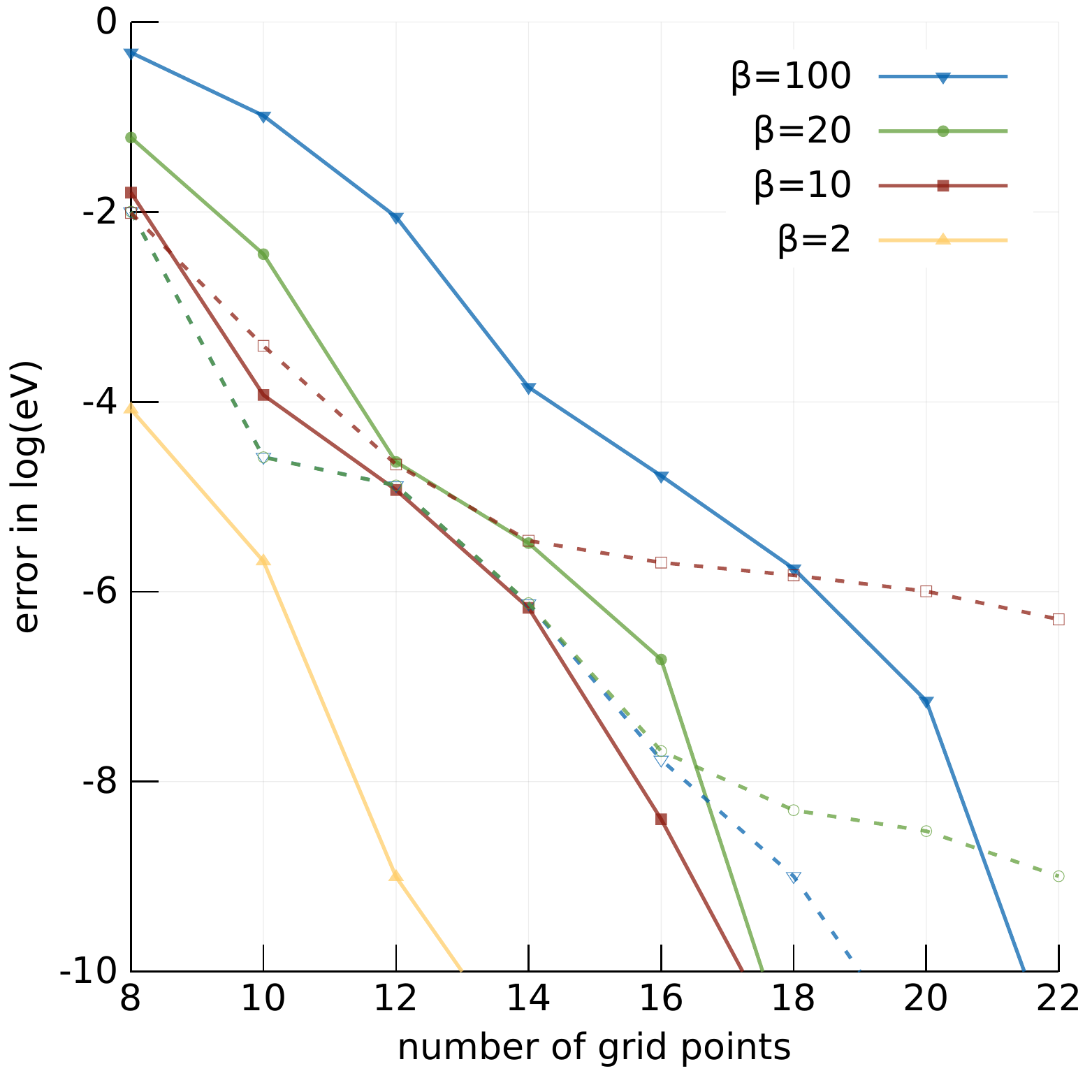}
\caption{
Grid convergence of RPA grand potential for Si at different temperatures (or
k-point smearings). Empty symbols correspond to  $\beta=\infty$ (zero temperature) implementation
using the same k-point smearing applied in the preceding Kohn-Sham groundstate
calculations. Inverse temperatures are in eV$^{-1}$.
}
\label{fig:SiRPA}
\end{figure}

We emphasize that in the limit $\beta\to\infty$ all basis functions approach the
IC basis functions used in the zero temperature RPA algorithms.\cite{Kaltak2014,
Helmich2016, Beuerle2018} That is, at $T$=0 K, the bosonic IA$_1$ and fermionic
IB$_1$ grid merge to the same frequency grid of IC$_1$. However, the zero- and
finite temperature grids can be compared only for systems with a finite band
gap, since using the $T$=0~K algorithm for metals results in problems and slow
convergence of the total energy with the number of grid points (the $L^2$-norm
of IC in Tab. \ref{tab:IsometricBases} diverges for $x\to0$). In contrast, the
IA$_1$-quadrature is valid for all systems (including metals) at all finite
temperatures. Hence, the Kohn-Luttinger conundrum\cite{PhysRev.118.41} is
circumvented, since the thermodynamic limit is performed at finite temperatures.
Consequently, a comparison of the grid convergence to our zero temperature
implementation of the RPA is useful only for systems with a finite band gap at
$T$=0 K, like for instance Si. The corresponding comparisons are given in
\fref{fig:SiRPA}.

\begin{figure}[tbp]
\centering
\MyFigure{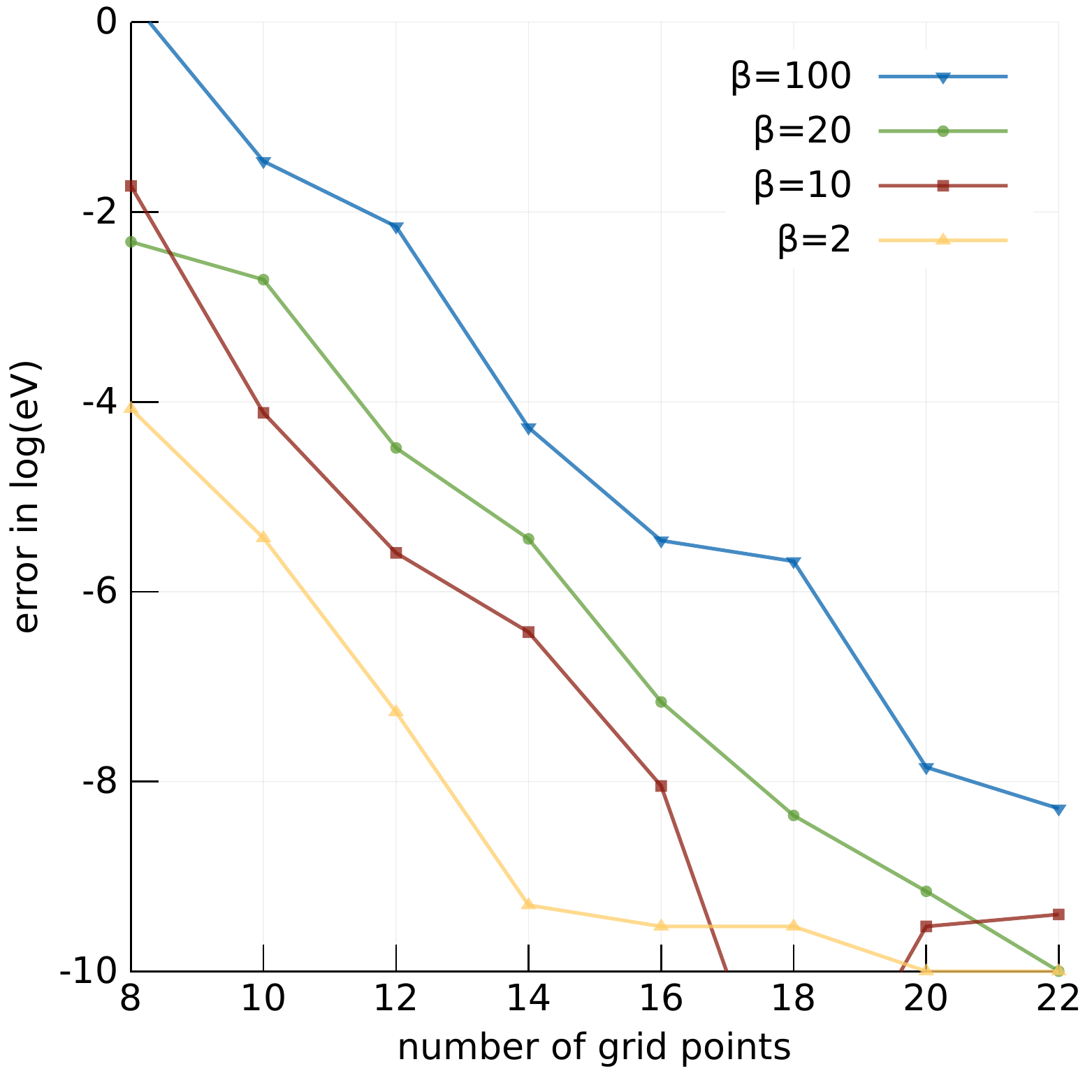}
\caption{
Grid convergence of RPA grand potential for SrVO$_3$ at
different inverse temperatures. Inverse temperatures are in eV$^{-1}$.
}
\label{fig:SrVO3RPA}
\end{figure}

The exponential grid convergence of the IA$_1$-quadrature for finite
temperatures is evident (solid lines).  The required number of points for a
given precision increases with decreasing temperature, because the minimization
interval increases linearly with $\beta$ and therefore the quadrature error
increases too.  Not surprisingly, a similar grid convergence rate is observed
for paramagnetic SrVO$_3$ as demonstrated in \fref{fig:SrVO3RPA}. This system is
known to be computationally challenging, because of the presence of several
degenerate, partially populated states around the chemical potential, even in
the limit $\beta\to\infty$. 

Comparing the IA$_1$-convergence rate with the zero temperature grid convergence
for Si, a similar slope is observed for $\beta=100$ eV$^{-1}$, see empty
triangles in \fref{fig:SiRPA}.  However, more IA$_1$-grid points for the same
precision as in the $T=0$ case are required. The zero temperature quadrature,
presented in another work of the authors,\cite{Kaltak2014} outperforms the
finite temperature grid at $\beta=100$ eV$^{-1}$ corresponding  to a sharp
k-point smearing of $\beta^{-1}=0.01$ eV. The reason is that the
zero temperature grid is "aware" of the band gap and designed to integrate that
as well as the largest excitation energies exactly. The finite temperature grid
is designed to work between 0 and the largest excitation energy (at a given
$\beta$).  As the temperature increases, partial occupancies are introduced.
This has the effect that the exponential convergence rate of the $T=0$ grid
deteriorates and causes the $T=0$ RPA algorithm even to converge towards a wrong
limit that differs from the finite temperature implementation (flattening of
dashed lines). Only for $\beta=100$ eV$^{-1}$ we observed that both, the zero-
and finite temperature RPA implementations, converge to the same result. This is
not surprising, because as $\beta$ becomes smaller, more states with energy
around $\epsilon_F$ become fractionally populated. These states are described
incorrectly by the zero temperature algorithm. Thus, we recommend to use the
finite temperature RPA algorithm for systems with a small or zero band gap. 

\subsection{ Performance of F-quadrature for $GW$ }
Finally, we have studied the grid convergence of the F-quadrature for Si and
paramagnetic SrVO$_3$ by calculating the GM grand potential in the $GW$
approximation using Equ. \eqref{eq:DefGMGrandPotential}. For demonstration
purposes, we have performed a single self-consistent update of the Green's
function (starting from the PBE Green's function), fixed the chemical potential
$\mu$ in the interacting Green's function \eqref{eq:DysonG} and self-energy to
the value of the non-interacting Green's function, and subsequently evaluated
the GM energy in the GW approximation. Tests using fully selfconsistent
calculations, indicate a similar convergence behavior. Fixing the chemical
potential means that the interacting Green's function for negative $\tau$
describes a system with a slightly different number of electrons $N_e$ in the
unit cell than the non-interacting counterpart.\footnote{ Fixing this requires
the adjustment of the chemical potential and re-calculation of the interacting
Green's function $G$ until $N_e=\lim_{\tau\to0-}\mathrm{Tr}G(-i\tau)$ is
satisfied. }
\begin{figure}[tbp]
\centering
\MyFigure{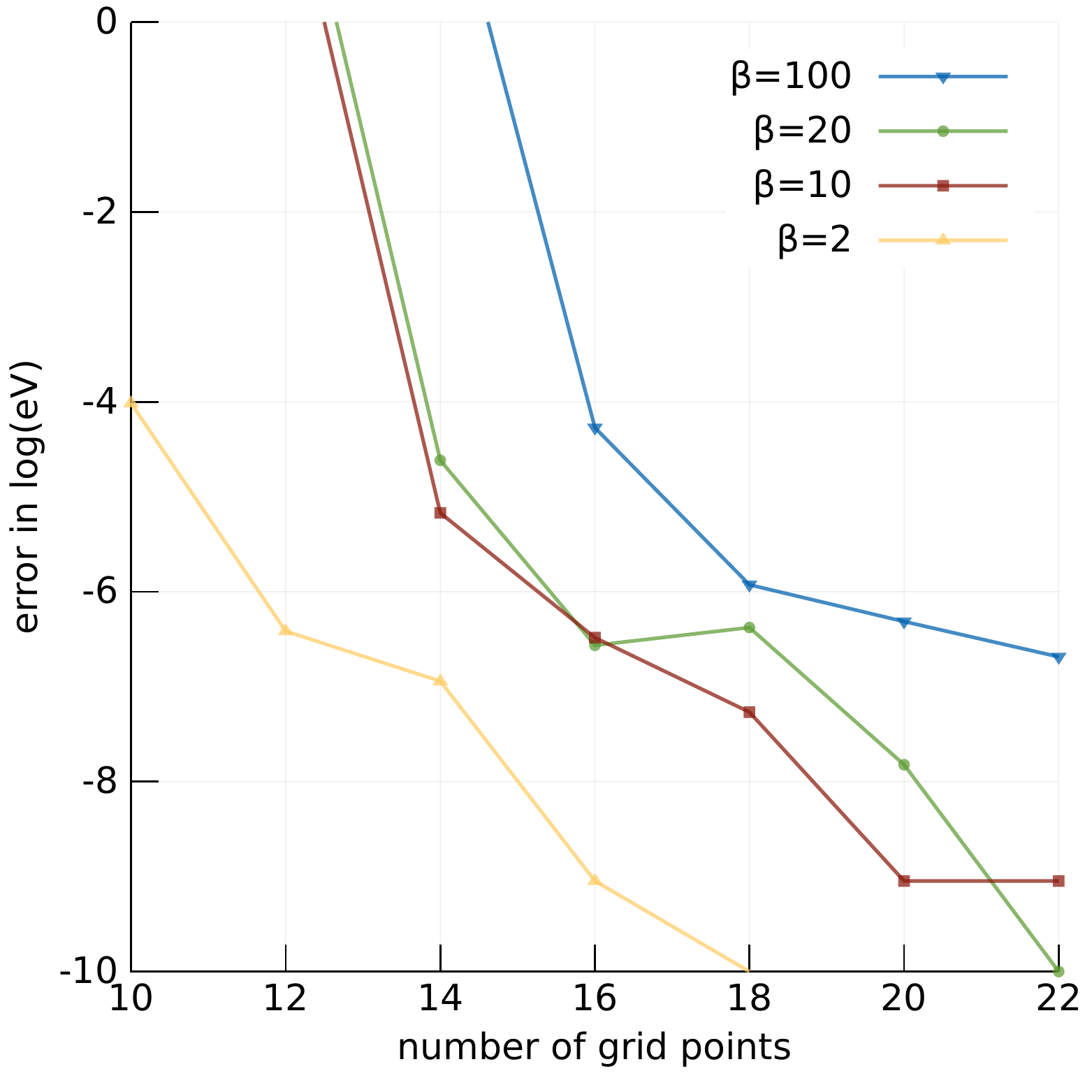}
\caption{F-grid convergence for GM grand potential Si. Inverse
temperatures are in eV$^{-1}$.}
\label{fig:GMSi}
\end{figure}

\begin{figure}[tbp]
\centering
\MyFigure{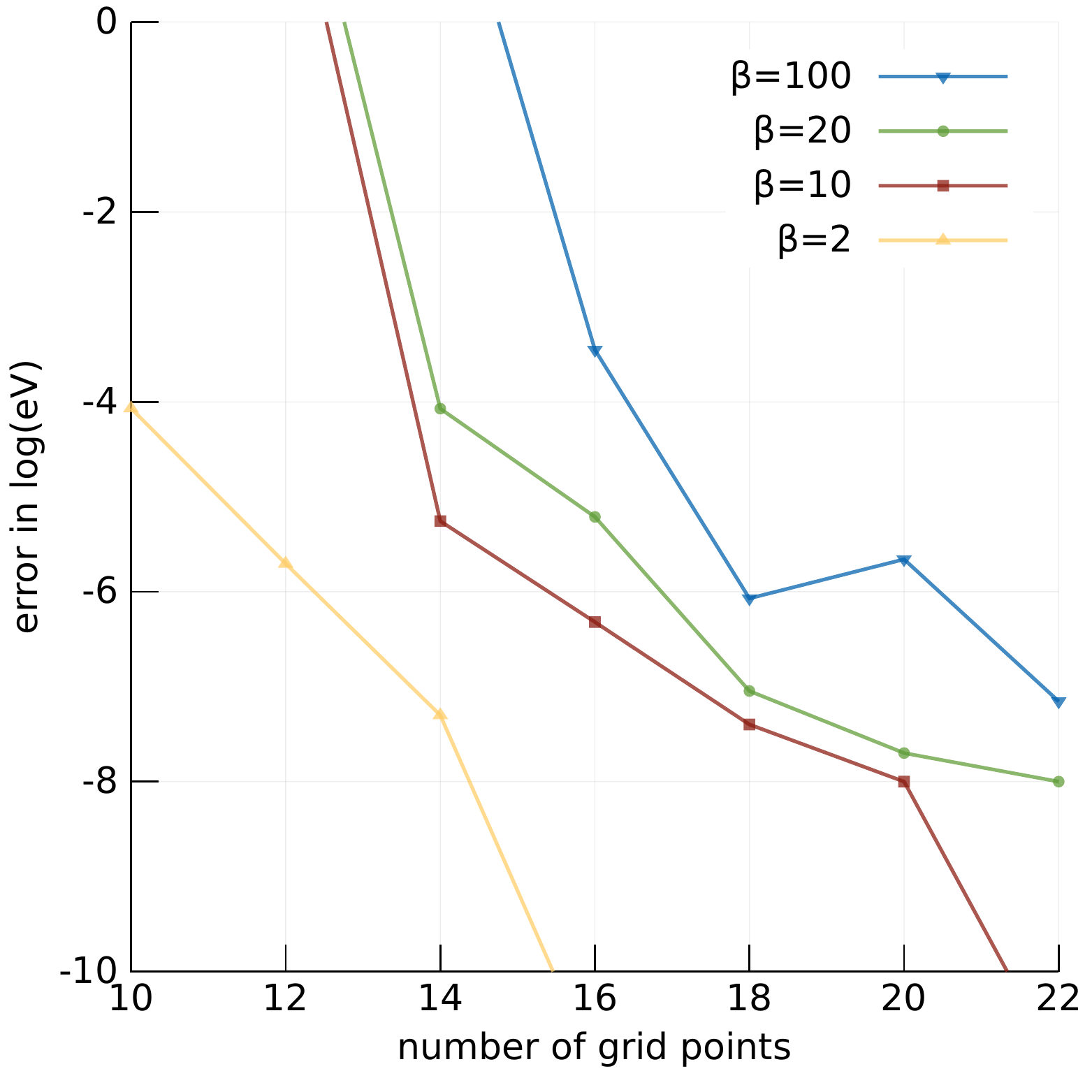}
\caption{F-grid convergence for GM grand potential SrVO$_3$.
Inverse temperatures are in eV$^{-1}$.}
\label{fig:GMSrVO3}
\end{figure}

The results for different values of $\beta$ of Si and SrVO$_3$ are given in
\fref{fig:GMSi} and \fref{fig:GMSrVO3}, respectively. One recognizes that the
grid convergence is very similar for both systems.  Nevertheless, the
convergence is worse compared to the RPA, because the F-quadrature error is
larger compared to the IA$_1$-error. 

However, our discussion in \ref{sec:MinimaxFourierTransformation} suggests that the present choice is the best compromise--- at least the best we could find ---,  
and necessitated by the need to have  the same time
grid for bosonic and fermionic functions, as well as a single frequency grid for fermionic functions. For practical applications, the error
of roughly 1 $\mu$eV  with 16 and more quadrature points is negligible. Other
convergence parameters, such as the energy cutoff of the basis set, typically
yield larger errors.\cite{PhysRevB.90.075125}

\begin{figure}[tbp]
\centering
\MyFigure{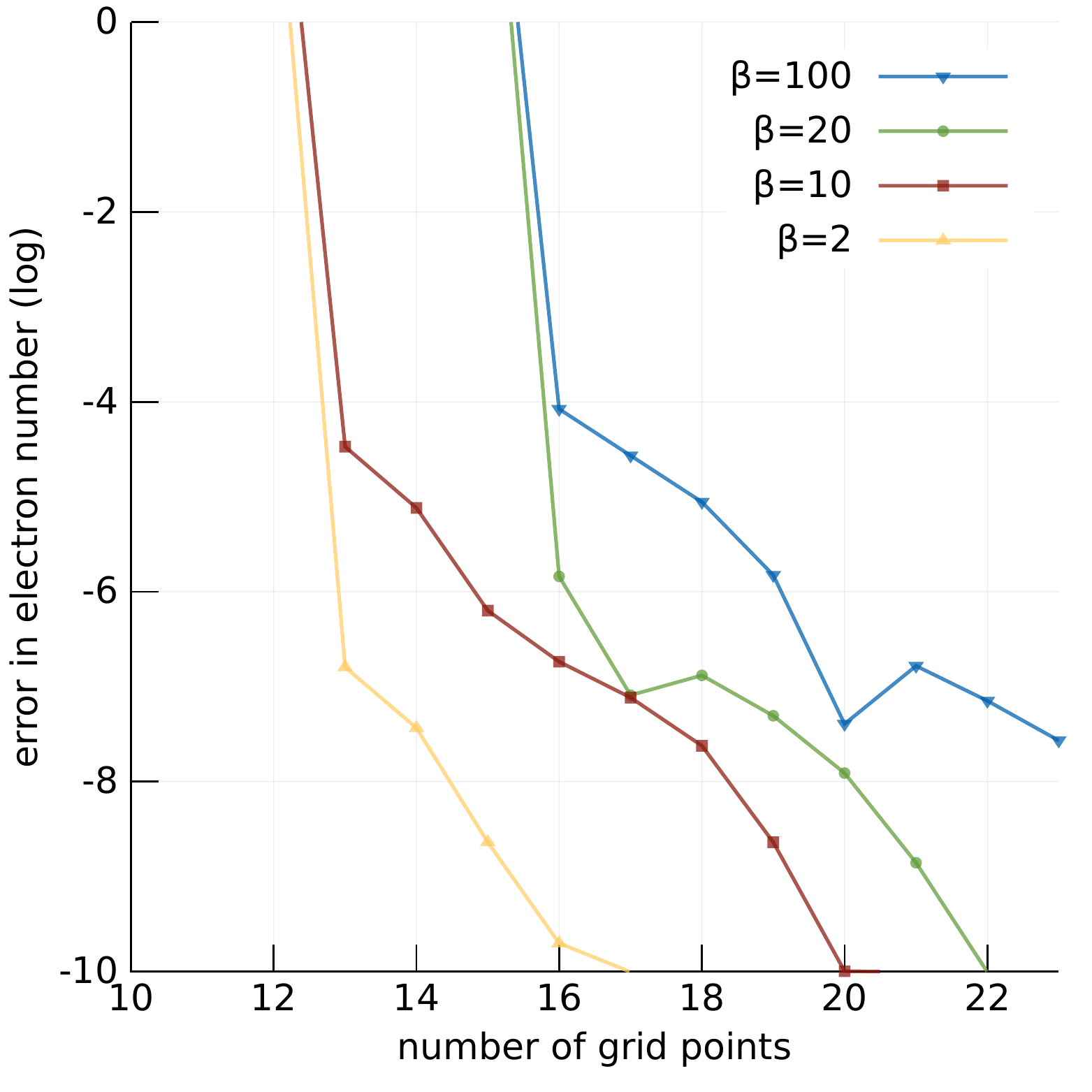}
\caption{Particle number conservation error $|N_e-N_e'|$ of F-grid when
calculating the electron density from the non-interacting Kohn-Sham propagator
of SrVO$_3$ (see text). Inverse temperatures are in eV$^{-1}$.}
\label{fig:ChargeConservationSrVO3}
\end{figure}

Last, we have considered the electron number conservation of the F-quadrature, that is
the difference of $|N_e-N'_e|$, where $N_e$ is the exact number of electrons in
the unit cell and $N'_e$ has been calculated from the trace of Equ. 
\eqref{eq:DensityMatrixQuadrature}. We have studied the non-interacting
propagator $\tilde g$ of Equ. \eqref{eq:DefG0} for SrVO$_3$. This corresponds to roughly
$1056\times64$ poles of the Green's function on the real-frequency axis in the
regime $|x|\le 400\beta$. The error in the particle number with the number of
F-quadrature points is shown in \fref{fig:ChargeConservationSrVO3}.
One can see that the convergence is exponential and increases and decreases
with $\beta$ in the same way as the RPA and GM energies. Not surprisingly, the
convergence is the same as compared to the case where the GM energy is used as
measure (see \fref{fig:GMSrVO3}). Also, the F-quadrature converges faster with
the number of grid points compared to the OHQ-quadrature (not shown). For
instance, the F-quadrature yields a precision of $10^{-10}$ states per unit cell
for $\beta=10$ using $N=20$ quadrature points, while the same precision is
only reached with $N=118$ OHQ-quadrature points.
\section{Conclusion}\label{sec:Conclusion}
We presented an efficient method for the Matsubara summation of bosonic and
fermionic correlation functions on the imaginary frequency axis. By constructing
optimum subspaces of the considered Hilbert space of dimension $N$, we obtained
imaginary time and frequency grids for all correlation functions appearing in
finite temperature perturbation theory.  Furthermore, using the argument of
$\varepsilon$-isometric spaces, we have shown that the transformation from
imaginary time to imaginary frequency can be performed with high precision. 

We implemented this technique in VASP to generalize our zero temperature random
phase approximation (RPA) and $GW$ algorithms to finite temperatures and
obtained a similar exponential grid convergence for the RPA grand potential (see
\fref{fig:SiRPA}) as in the $T=0$ case.\cite{Kaltak2014} To reach
$\mu$eV-accuracy, typically, less than 20 grid points are required. This holds
true even for low temperatures, so that the RPA grand potential can be evaluated
very efficiently for insulating as well as metallic systems with a computational
complexity that grows only cubically with the number of electrons in the unit
cell. 

Furthermore, we showed how to choose the frequency grid for fermionic
correlation functions and how to evaluate the Galitskii-Migdal grand potential
at finite temperatures using the F-quadrature (see sections
\ref{sec:FermionicFrequencyGrid} and \ref{sec:MinimaxFourierTransformation}).
Here a compromise between $\varepsilon$-isometry and integration efficiency has
to be made that deteriorates the grid convergence slightly compared to the RPA.
For practical applications, however, the precision of the Matsubara summation is
still sufficiently good. Other error sources, such as basis set errors will
usually dominate.

In summary, we showed that optimized grids can be found for the accurate
Matsubara summation of both, bosonic and fermionic functions, with roughly 20
grid points. The hypergeometric grids of Ozaki\cite{Ozaki2007} and Hu {\em et.
al.}\cite{Hu2010} (see section \ref{sec:MinimaxIsometry}) require roughly 100
and more points for the same precision at low temperatures.
\begin{appendix}

\section{Second order contribution to the correlation energy at finite $T$}\label{app:Proof2ndorder}

In this appendix, we show that the conserved $L^2$-norm
of the IA isometry describes the second order contribution to the
correlation part of the grand-canonical potential defined in Equ.
\eqref{eq:DefMP2GrandPotential} at finite temperature. We prove this for $\beta=1$ and
the diagonal matrix elements $\Omega_{\alpha\gamma}=\sum_{n\in\BoldMath{Z}} \tilde \chi^2_{\alpha\gamma}(i\nu_n)$ using the explicit form for the
polarizability in imaginary frequency
\begin{equation}\label{eq:ChiExplict}
\begin{split}
\tilde\chi^2_{\alpha\gamma}(i\nu_n) =&\left[ \frac12 \frac{\Delta_{\alpha\gamma}}{
\Delta_{\alpha\gamma}^2+\nu_n^2}\right]^2
\left[f(x_\alpha)-f(x_\gamma)\right]^2\\
\Delta_{\alpha\gamma}=&x_\alpha-x_\gamma.
\end{split}
\end{equation}
From the series representation of the hyperbolic cotangent
\eqref{eq:CothSeries}, and the identity
\begin{equation}\label{eq:qAnalog}
\frac{1}{\left(\Delta_{\alpha\gamma}^2+\nu_n^2\right)^2} = \lim_{q\to0}
\frac{1}{2q}\left[\frac{1}{\Delta_{\alpha\gamma}^2+\nu_n^2-q } 
-\frac{1}{\Delta_{\alpha\gamma}^2+\nu_n^2+q }\right], 
\end{equation}
and the addition theorem for the hyperbolic tangent\cite{abramowitzstegun}
\begin{equation}
\label{eq:AdditionTheoremTanh}
\tanh\frac{\Delta_{\alpha\gamma}}{2}
=
\frac{\tanh\frac{x_{\alpha}}{2}
-\tanh\frac{x_{\gamma}}{2}}{
1-\tanh\frac{x_{\alpha}}{2}
\tanh\frac{x_{\gamma}}{2}
}\,,
\end{equation}
it is easy to show that 
\begin{equation}
\label{eq:MP2Contribution}
\begin{split}
\Omega_{\alpha\gamma}=&\frac14
\left[ \frac{f(x_\alpha)-f(x_\gamma)}{\tanh\frac{\Delta_{\alpha\gamma}}{2}} \right]^2
\\
&\times\left[
\frac18
\left(1-\tanh^2\frac{\Delta_{\alpha\gamma}}{2}\right)
+\frac14
\frac{
\tanh^2\frac{\Delta_{\alpha\gamma}}{2}}{\Delta_{\alpha\gamma} }
\right].
\end{split}
\end{equation}
For the sake of simplicity, the Coulomb matrix elements have been suppressed. 
The last factor in this expression corresponds to the conserved $L^2$-norm of
the IA$_1$ isometry in Tab. \ref{tab:IsometricBases}, while the first factor
is non-zero for all values of $\Delta_{\alpha\gamma}$ so that the 
identity $\sum_{n\in\BoldMath{Z}} \tilde \chi^2_{\alpha\gamma}(i\nu_n) =
\Omega_{\alpha\gamma}$ can be divided by the same factor proving our assertion.
The proof can be generalized to the off-diagonal elements as well.

\section{Why frequency grids for GM are difficult}
\label{sec:GMdifficult}
 First, we consider the frequency dependence of the free propagator
\eqref{eq:DefG0}. The cosine and sine transformations of the odd and even time
basis functions \eqref{eq:BasisTimeOdd} and \eqref{eq:BasisTimeEven} for
fermionic frequencies \eqref{eq:FermionicFrequencies} are given in Eqs.
\eqref{eq:SymmetricPartG0InFreq} and \eqref{eq:AntisymmetricPartG0InFreq},
respectively. 
Then the non-interacting propagator \eqref{eq:DefG0} on the fermionic Matsubara
axis reads 
\begin{equation}
\label{eq:G0InFrequency}
\tilde g(x_\alpha,i\omega_m) = \tilde u_{\omega_m}(x_\alpha)+ i \tilde
v_{\omega_m}(x_\alpha).
\end{equation}
Second, we observe that every fermionic function can be decomposed into terms
that are even and odd in $\omega$, including the product of the propagator and
self-energy as it appears in the GM grand potential
\eqref{eq:DefGMGrandPotential}. It is, obvious, that only the real part of the
product $\tilde G\tilde \Sigma$ contributes to the total energy. Thus the most
general matrix element, which gives a non-zero contribution to the GM grand
potential has the form\footnote{The imaginary part is proportional to odd terms
in the frequency 
$\tilde u_m(\epsilon_\alpha)\tilde v_m(\delta_\gamma)+\tilde
v_m(\epsilon_\alpha)\tilde u_m(\delta_\gamma)$
and vanishes when the sum over all (positive and negative) fermionic Matsubara
frequencies is carried out.  }
\begin{equation}\label{eq:GMMatrixElement}
\begin{split}
\tilde G(i\omega_m)\tilde\Sigma(i\omega_m)=&\sum\limits_{-x_{\max}\le
x,y\le x_{\max}}D_{G}(x)D_\Sigma(y)\\
\times&\left[\tilde u_{\omega_m}(x)\tilde u_{\omega_m}(y) - 
\tilde v_{\omega_m}(x)\tilde v_{\omega_m}(y)\right],
\end{split}
\end{equation}
where $x$ and $y$ are the poles of the Green's function and the self-energy on
the real-frequency axis and $D_G,D_\Sigma$ the spectral densities, respectively.
Without loss of generality, we set $D_G=D_\Sigma=1$ and assume that the magnitudes of the poles
are smaller than a positive number, i.e. $|x|,|y|\le x_{\max}$.
Third, we note that the analogue of the IA$_1$-quadrature of bosonic functions
\eqref{eq:MinimaxProblemA1} for fermionic ones
\begin{equation}\label{eq:MinimaxProblemFerF2}
\min_{\sigma_k>0,\omega_k\in(0,\infty)}
\max_{0\le x\le x_{\rm max}}
\left|
\|x\|^2_2 - \sum\limits_{k=1}^{N}\sigma_k \tilde u^2_k(x)
\right|
\end{equation}
yields the IB$_1$-quadrature, see Tab. \ref{tab:IsometricBases}. Unfortunately,
the IB$_1$-quadrature only allows to evaluate the first term on the right hand side of
\eqref{eq:GMMatrixElement} accurately, but fails for the product of two odd
functions $\tilde v$. Similarly, the IA$_2$-quadrature obtained from the minimax
problem
\begin{equation}\label{eq:MinimaxProblemFerA2}
\min_{\sigma_k>0,\omega_k\in(0,\infty)}
\max_{0\le x\le x_{\rm max}}
\left|
\|x\|^2_2 - \sum\limits_{k=1}^{N}\sigma_k \tilde v^2_k(x)
\right|
\end{equation}
that approximates the same norm as the time and IA$_1$-quadrature, describes 
only the second term in \eqref{eq:GMMatrixElement}. Consequently, neither the
IB$_1$- nor the IA$_2$-quadrature can be used for our purposes.  

\section{Poisson summation and hyperbolic functions: A proof of Equ.
\eqref{eq:L1NormDef}}\label{app:ProofTanh}
To proof identity \eqref{eq:L1NormDef}, we use Poissons summation
formula\cite{Higgins1985}
\begin{equation}\label{eq:Poisson} 
\sum\limits_{n\in\BoldMath{Z}}f(n) = 
\sum\limits_{k\in\BoldMath{Z}}\tilde f(k) 
\end{equation}
for a function $f$ and its Fourier transform $\tilde f$.  Inserting $f(tz)
= e^{-2|tz|}$ into the left hand side of \eqref{eq:Poisson} one obtains with the
geometric series of the hyperbolic cotangent 
\begin{equation}\label{eq:IdentityCoth}
\sum\limits_{n\in\BoldMath{Z}}e^{-2|n z|} = 
\frac{1+e^{-2|z|}}{1-e^{-2|z|}}=\coth|z|.
\end{equation}
Consequently, evaluating the Fourier integral gives 
\begin{equation}\label{eq:FourierIntegral}
\tilde f(k)= \int_{-\infty}^\infty\mathrm{d}t e^{-2|tz|}
 e^{i 2\pi k t}=\frac{|z|}{\pi^2 k^2+|z|^2},
\end{equation}
which after inserting into the right hand side of \eqref{eq:Poisson} yields the
identity
\begin{equation}\label{eq:CothSeriesZ}
\coth|z|=\sum\limits_{k\in\BoldMath{Z}}\frac{|z|}{\pi^2 k^2+|z|^2}.
\end{equation}
On the one hand, replacing $|z|\to |z|/2$ and dividing by $2$, this identity
becomes
\begin{equation}\label{eq:CothSeries}
\frac12\coth\frac{|z|}2=\sum\limits_{k\in\BoldMath{Z}}\frac{|z|}{(\pi 2k)^2+|z|^2}.
\end{equation}
On the other hand, the series \eqref{eq:CothSeriesZ} on the right hand side can
be split into a series over even and a series over odd integers
\begin{equation}\label{eq:SeriesCothSplit}
\coth|z|=\underbrace{\sum\limits_{k\in\BoldMath{Z}}\frac{|z|}{(\pi2k)^2+|z|^2}}_{=\frac12\coth\frac{|z|}2}
+\sum\limits_{k\in\BoldMath{Z}}\frac{|z|}{(\pi(2k+1))^2+|z|^2} 
\end{equation}
The first term on the right hand side follows from \eqref{eq:CothSeries}, while
the second part is the left hand side of Equ. \eqref{eq:L1NormDef}. After
comparison with the well-known hyperbolic identity
\begin{equation}\label{eq:HyperbolicIdentity}
\coth z = \frac12\coth\frac{z}2 + \frac12\tanh\frac{z}2,
\end{equation}
one identifies the second term in \eqref{eq:SeriesCothSplit} with 
\begin{equation}\label{eq:TanhSeries}
\frac12\tanh\frac{|z|}2 
=\sum\limits_{k\in\BoldMath{Z}}\frac{|z|}{(\pi(2k+1))^2+|z|^2}
\end{equation}
and Equ. \eqref{eq:L1NormDef} is proven. 

Note, the derivative of the left and right hand side of Eqs.
\eqref{eq:TanhSeries} and \eqref{eq:CothSeries} in combination with the q-analog
\eqref{eq:qAnalog} gives an alternative way to calculate the $L^2$-norms
tabulated in Tab.  \ref{tab:IsometricBases}. 
\end{appendix}

\bibliography{MyReferences}

\end{document}